\documentclass[final,5p,times,twocolumn]{elsarticle}
\usepackage{listings}
\usepackage{xcolor}
\usepackage{amsmath} 
\usepackage[bottom]{footmisc}

\usepackage{caption}
\usepackage{subcaption}

\definecolor{bgcolor}{RGB}{248, 248, 248} 
\definecolor{keywordcolor}{RGB}{0, 92, 97} 
\definecolor{stringcolor}{RGB}{209, 154, 102} 
\definecolor{commentcolor}{RGB}{106, 115, 125} 
\definecolor{numbercolor}{RGB}{9, 134, 88} 
\definecolor{framecolor}{RGB}{200, 200, 200} 

\lstdefinestyle{modern}{
    backgroundcolor=\color{bgcolor},   
    keywordstyle=\color{keywordcolor}\bfseries, 
    stringstyle=\color{stringcolor}, 
    commentstyle=\color{commentcolor}\itshape, 
    numberstyle=\color{numbercolor}, 
    basicstyle=\ttfamily\footnotesize, 
    frame=single, 
    rulecolor=\color{gray}, 
    breaklines=true, 
    tabsize=4, 
    numbers=left, 
    stepnumber=1, 
    showstringspaces=false, 
    xleftmargin=10pt,
}
\usepackage{booktabs} 
\usepackage{graphicx}
\usepackage{multicol} 
\usepackage{footnote}

\usepackage{hyperref}

\journal{Future Generation Computer Systems}

\begin{document}

\begin{frontmatter}



\title{Offloading Artificial Intelligence Workloads across the Computing Continuum by means of Active Storage Systems}

\author[bsc]{Alex Barcel\'o}
\ead{alex.barcelo@bsc.es}
\affiliation[bsc]{organization={Barcelona Supercomputing Center (BSC)}, city={Barcelona}}
\author[ceadar]{Sebastián A. Cajas Ordoñez}
\ead{sebastian.cajasordonez@ucd.ie}
\author[ceadar]{Jaydeep Samanta}
\ead{jaydeep.samanta@ucd.ie}
\author[ceadar]{Andrés L. Suárez-Cetrulo}
\ead{andres.suarez-cetrulo@ucd.ie}
\author[ceadar]{Romila Ghosh}
\ead{romila.ghosh@ucd.ie}
\author[ceadar]{Ricardo Simón Carbajo}
\ead{ricardo.simoncarbajo@ucd.ie}
\affiliation[ceadar]{organization={Ireland’s Centre for Artificial Intelligence (CeADAR)}, city={Dublin}}
\author[upc]{Anna Queralt}
\ead{anna.queralt@upc.edu}
\affiliation[upc]{organization={Universitat Politècnica de Catalunya (UPC)}, city={Barcelona}}


\begin{abstract}
The increasing demand for artificial intelligence (AI) workloads across diverse computing environments has driven the need for more efficient data management strategies. Traditional cloud-based architectures struggle to handle the sheer volume and velocity of AI-driven data, leading to inefficiencies in storage, computation, and data movement. This paper explores the integration of active storage systems within the computing continuum to optimize AI workload distribution.

By embedding computation directly into storage architectures, active storage is able to reduce data transfer overhead, enhancing performance and improving resource utilization. Other existing frameworks and architectures offer mechanisms to distribute certain AI processes across distributed environments; however, they lack the flexibility and adaptability that the continuum requires, both regarding the heterogeneity of devices and the rapid-changing algorithms and models being used by domain experts and researchers.

This article proposes a software architecture aimed at seamlessly distributing AI workloads across the computing continuum, and presents its implementation using mainstream Python libraries and dataClay, an active storage platform. The evaluation shows the benefits and trade-offs regarding memory consumption, storage requirements, training times, and execution efficiency across different devices. Experimental results demonstrate that the process of offloading workloads through active storage significantly improves memory efficiency and training speeds while maintaining accuracy. Our findings highlight the potential of active storage to revolutionize AI workload management, making distributed AI deployments more scalable and resource-efficient with a very low entry barrier for domain experts and application developers.
\end{abstract}


\end{frontmatter}




\section{Introduction}
\label{sec:introduction}

Artificial Intelligence (AI) plays an important role in modern society, revolutionizing industries and reshaping scientific research. A wide variety of fields are driving research and adoption of artificial intelligence: healthcare and medicine \cite{rong2020artificial, bhinder2021artificial}, marketing \cite{verma2021artificial}, power and grid \cite{raza2015review}, robotics \cite{soori2023artificial}, and many others. AI continues to evolve, and we see how its influence extends across disciplines, driving both theoretical breakthroughs and practical advancements.

Computing continuum environments (environments combining IoT devices, edge devices, and cloud computing) can greatly benefit from the integration of AI applications \cite{wang2020convergence, deng2020edge, ICOS-paper}. These environments generate vast amounts of data at distributed locations, where real-time processing and decision-making are often critical. By deploying AI closer to the data sources –-within edge devices instead of in the cloud–- the system achieves a better response,  operating efficiently even in bandwidth-constrained or latency-sensitive scenarios \cite{shi2016edge}. However, leveraging the edge in the proper way is something that requires care; an unbalanced infrastructure and load can result in performance inversion, as discussed by Ali-Eldin \emph{et al.} in \cite{ali2021hidden}. This shows how important and impactful software solutions are in computing continuum architectures, and how these solutions need to be able to perform data management and task distribution in a malleable manner.

Despite the potential advantages of integrating AI into computing continuum environments, several challenges arise in terms of data management and execution placement. One key challenge is handling the vast and heterogeneous data generated by IoT and edge devices. Efficiently collecting, processing, and storing this data while ensuring quality, consistency, and security is complex, especially in distributed settings with limited resources \cite{ICOS-paper}. Additionally, placing AI execution close to the data and where results are needed requires balancing computational workload across edge and cloud infrastructures. Resource-constrained edge devices may struggle with the high computational demands of AI models, necessitating optimization techniques or dynamic workload distribution \cite{ordonezadaptive}. Ensuring seamless coordination between different layers of the continuum while maintaining low latency, energy efficiency, and data privacy remains an ongoing research challenge.

Technological solutions that effectively combine data management requirements with execution constraints in computing continuum environments are scarce. Many existing AI frameworks and workflow managers are designed either for cloud-scale processing \cite{sagemaker-web, vertexai-web, ray-web, metaflow-web}, where resources are abundant, or for lightweight edge deployments \cite{litert-web}, where computational power is limited. However, seamless integration across those paradigms remains an open challenge \cite{mukhopadhyay2019heterogeneous,xu2020edge}. This article proposes an architecture that can be used in such computing continuum environments, spanning from the edge to the cloud, and is capable of leveraging heterogeneous devices and using their resources seamlessly. Having mechanisms to dynamically offload workloads is key to handling data flows. Currently, the lack of standardized, end-to-end, flexible solutions makes it difficult for developers to deploy AI applications that can be efficiently adapted and operate across IoT, edge, and cloud environments.

The \emph{active} architecture approach to storage systems can help address these challenges by enabling more intelligent data management and processing within the computing continuum. The origin of the term \emph{active} comes from the database world, from the SAMOS\cite{gatziu1992samos} project in 1992. However, it can be extended to other storage systems. An example of an \emph{active} object store (and also the one that we will be focusing on in the evaluation) is dataClay\cite{marti2017dataclay}. Generally, \emph{active} systems are able to seamlessly perform execution next to their data. Within the continuum, this feature can be leveraged to offload execution and run distributed workloads. This can be achieved without requiring any changes in the programming model of the application; an AI workload using an \emph{active} object store is capable of triggering the execution of its workload into specific devices (those where data resides). This is an invaluable tool for optimizing resource utilization and reducing data movement overhead across the computing continuum.

This article demonstrates the potential of active storage systems in distributing AI workloads across the computing continuum. To achieve this objective, we propose an architecture that leverages active storage systems. These systems have the potential to perform task offloading for AI workloads. The process of offloading tasks is a flexible one that can be adapted to different kinds of workloads and can take advantage of the heterogeneity of computing resources --catering to a diverse range of hardware in terms of cost, performance, resources, etc. The proposed objective can be reached without requiring the programming language to be redefined while maintaining the accuracy of AI results. The architecture we propose is an invaluable asset for defining scalable and distributed AI workloads across the computing continuum.

The key contributions of our proposal are the following:
\begin{itemize}
    \item A software architecture designed to optimize AI workloads in the computing continuum. The proposed architecture is based on the active storage approach, reducing data transfers and improving performance and resource utilization.
    \item An implementation of the proposed architecture using dataClay, an open-source active storage system. This implementation shows that the proposed architecture can seamlessly take advantage of heterogeneous resources across the continuum, supports mainstream AI Python libraries, and requires no changes at the programming paradigm level.
    \item An extensive evaluation demonstrating that our solution significantly improves training speeds while reducing the requirements of the most constrained devices in the continuum, maintaining accuracy of results at the same time.
\end{itemize}

This article is structured as follows. Section~\ref{sec:relwork} surveys state-of-the-art technologies, including hardware and software frameworks, that support distributed AI across the computing continuum. Section~\ref{sec:arch} contains our proposed architecture and outlines the role of the active storage system and technical details regarding the software stack that satisfies this architecture. Section~\ref{sec:methodology} describes the methodology used in the evaluation, detailing the workload characteristics and the experimental setups. Section~\ref{sec:results} presents the experimental results and analyzes them in comparison with the baseline (non-distributed experiments). Section~\ref{sec:towards-distributed} validates the scalability of the storage system for distributed workloads. Section~\ref{sec:reflections} elaborates on the limitations, the full potential, and the future work related to our proposal. Finally, Section~\ref{sec:conclusions} draws the main conclusions of the article.

\section{Related Work}
\label{sec:relwork}

The quantity of data and the size of AI models used in HPC environments have been increasing in recent years. For instance, given the huge data requirements of Deep Learning, there has been some focus on the I/O aspect \cite{aizman2019high} as well as the computing distribution \cite{ejarque2022enabling}. Although both HPC and computing continuum infrastructures are highly distributed and constituted by lots of devices, the HPC ecosystem is centralized, homogeneous, and well-connected, while computing continuum infrastructures are the opposite. However, here is a common goal in both scenarios: taking advantage of distributed resources and using them efficiently.

Being able to run ML and AI in smaller, power-efficient devices is a pursuit that can be addressed from the hardware point of view or from the software point of view. From the hardware point of view, new power-efficient TPU devices (such as the Coral edge TPU\cite{coral} among others\cite{seshadri2022evaluation}) allow power-efficient workload execution on constrained devices by efficiently using specialized hardware. Software efforts have produced algorithms and models that are suitable for constrained environments \cite{ordonezadaptive} and frameworks adapted to run ``on-device AI'' (i.e., not in the cloud but in constrained devices) such as LiteRT \cite{litert-web}. These joint efforts render it feasible to run ``moderately small'' workloads on the edge \cite{ray2022review}. Finding efficient workload offloading mechanisms while exploiting data locality is still a research challenge that we aim to tackle in this paper, something that is possible thanks to all the effort from this field but has not yet been solved by prior work.

The ability to run intelligence workloads across different devices with different characteristics is a key aspect of any continuum deployment, and as such, it has been the subject of multiple studies. 
For instance, Rosendo \emph{et al.}\cite{rosendo2022distributed} reviews the different alternatives and impact of the libraries and frameworks that can be used to distribute such workloads. Their work identifies the challenges and requirements needed to enable intelligence across the Edge-to-Cloud Continuum in an efficient way. Our goal is to provide an architecture that leverages task offloading, taking advantage of the distributed edge-to-cloud resources in the infrastructure.

The need to distribute the execution of AI workloads has been the object of study for a while now. For instance, BentoML \cite{bentoml-repo} is a framework that aims to solve the issue of distributing and scaling AI inference, with options for offloading through Kubernetes. Although it is envisioned for the cloud (``any model, on any cloud''), it is perfectly usable for heterogeneous infrastructures like the computing continuum. BentoML is a valuable framework, but its main focus is model serving through APIs; this framework lacks Python programmability. Tools such as SageMaker \cite{sagemaker-web}, do not provide native orchestration for the edge, nor do they offer functionality to exploit data locality.

Other frameworks in the distributed AI landscape primarily focus on MLOps tasks such as model serving, pipeline orchestration, deployment automation, and continuous integration. These include MLFlow \cite{Zaharia2018AcceleratingMLflow}, Kubeflow \cite{Bisong2019KubeflowPipelines}, MLReef \cite{KreuzbergerMachineArchitecture}, ZenML \cite{Bodor2023MLOps:Directions}, MLRun \cite{Chaves2023TheComparative}, CML \cite{Hewage2022MachineSupport}, Seldon Core \cite{Klaise2020MonitoringProduction}, and Yatai \cite{yatai-web}. However, these solutions target operational pipelines and are primarily cloud-native, relying on assumptions like centralized resources and stable connectivity. This makes them unsuitable for the heterogeneous, resource-constrained, and intermittently connected environments typical of the continuum due to container overhead and a lack of mechanisms to leverage localized data.

Another popular AI tool focused on distributed environments is Flower \cite{beutel2020flower}, a framework that offers a library-agnostic approach to perform federated learning, and it is suitable for the computing continuum. While the federated learning workloads run by Flower are indeed very relevant in the recent AI literature to achieve privacy-aware learning, they are neither a solution nor a complete picture for every distributed AI scenario. The solution we propose is a complementary tool that addresses data management and task distribution. Later, in Section~\ref{sec:reflections}, we further discuss the synergy and dynamics that may appear when combining Flower with our solution.

The active feature for storage systems has been explored and implemented on several object stores. For instance, the distributed object store Ceph RADOS \cite{weil2007rados} features the concept of \emph{object classes}~\cite{fisk2017mastering}, a feature that adds (limited) active mechanisms to its objects. Similarly,  Storlets~\cite{openstack-storlets}, which belongs to OpenStack, adds active features to Swift \cite{swift-web} objects. However, these object stores are entirely focused on the cloud and are not suitable for heterogeneous, highly decentralized continuum infrastructures. The work of Lofstead \emph{et al.}~\cite{lofstead2016daos} proposes an architecture with \emph{Function and Analysis Shipping} (a kind of active mechanism) using DAOS \cite{daos-web} as backend. The focus of this last storage system is on HPC environments, and it is not a flexible solution applicable to other infrastructures. 

One of the pillars of the active storage system architecture that we propose is the object storage (understanding \emph{object} as in the object-oriented paradigm). Besides dataClay (the active object store that we use in our proposed architecture), there are other storage and object-oriented programming paradigms that make use of this strategy. From the point of view of parallel and distributed execution, Charm++ \cite{kale1993charm++} is an established parallel programming framework that includes the programming model and the execution model. Charm++ is based on C++ (an object-oriented programming language) and it demonstrates the potential and programmability that working with distributed \emph{objects} brings to the table. Built on top of Charm++ is CharmPy \cite{galvez2018charmpy}, a similar parallel programming model for Python. However, the Charm++ architecture focuses on massive parallel applications and HPC environments, and it lacks the flexibility that we need for the computing continuum. The programming model behind Charm++ is rich and powerful, but it presents a steep entry level when trying to use it in the Artificial Intelligence domain, due to its lack of support to the applications and libraries commonly used in the AI ecosystem. Other object databases exist, such as ZODB \cite{zodb-web}. ZODB offers a natural Python interface and is closer to the Python programming language and to Python objects, which results in easier migration and easier integration with Python AI libraries. However, ZODB does not offer flexible distributed primitives and its deployment in a continuum environment is not straightforward. It does offer a specific mechanism for cross-database referencing, which can constitute a first step towards distributed scenarios, but its general architecture is based on centralized databases and the system is more suited for cloud-centric environments.

Serverless and Function-as-a-Service (FaaS) architectures offer an efficient, event-driven execution model that abstracts infrastructure management, providing scalability, fault tolerance, and ease of integration. There are multiple on-premises solutions that offer these features, such as OpenFaaS \cite{openfaas-web}, Knative \cite{knative-web}, or Apache OpenWhisk \cite{openwhisk-web}. However, these solutions are neither designed for heterogeneous infrastructures nor do they include data management. Their deployments rely mainly on underlying orchestrators (Kubernetes, Apache MESOS, Docker Compose\ldots) and it is not feasible to run them efficiently across the computing continuum with all its network and hardware variety. Our proposal shares key advantages such as infrastructure abstraction and ease of integration. Unlike these preexisting solutions, it also accommodates the heterogeneity of the computing continuum and provides data management capabilities. Merlino \emph{et al.} work \cite{merlino2024faas} demonstrates a promising architecture to enable the Function-as-a-Service model within the computing continuum paradigm. That work shows the relevance of modern computing continuum infrastructures and work distribution. Our proposal focuses on data management and the programming model, resulting in a low entry barrier for domain experts.

Most cloud providers include Function-as-a-Service among their offerings (AWS Lambda from Amazon, Azure Functions from Microsoft, Cloud Run functions from Google, etc.), which can be valuable for certain IoT deployments. However, using a cloud provider for data processing requires to transmit everything to the cloud and process it there, which results in extra costs and vendor lock-in. Taking advantage of the computing continuum is a key strategy for achieving scalability, data sovereignty, and efficiency. Amazon S3 Object Lambda \cite{s3objectlambda-web} attempts to bring computation closer to data by enabling in-situ processing during object operations (store, query, retrieval; that is GET, HEAD, and LIST requests). While this is conceptually aligned with our data management architecture, it remains confined within the cloud provider’s storage and execution ecosystem. As a result, it cannot support continuum-wide deployments across heterogeneous and resource-constrained environments.


In the field of computing continuum, Tallent et al. work \cite{tallent2024final} aims to optimize workflows via advanced measurement and compression techniques. Ferrer et al. \cite{ferrer2021towards} introduced cognitive continuum frameworks enabling autonomous decisions using local resources. Marru et al. \cite{marru2023cybershuttle} presented CyberShuttle for decentralized automated data management, complemented by Balouek-Thomert et al. \cite{balouek2019towards}, who developed adaptive edge-to-cloud data movement strategies. All these prior works illustrate the interest and efforts related to data management in computing continuum environments. The novelty in our proposal lies in the way we address the challenges of data distribution on the continuum. Our proposal aims to bring computation close to its data (through an active storage system) while maintaining the original programming framework and offering a transparent interface to the domain expert programming the application.



\section{Architecture}
\label{sec:arch}

\begin{figure}
    \centering
    \includegraphics[scale=0.45]{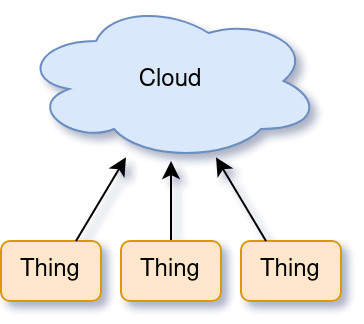}
    \caption{Traditional architecture of \emph{Things} and cloud}
    \label{fig:arch-trad}
\end{figure}

The historical starting point of the \emph{Internet of Things} was based on light devices (the \emph{things}) that sent information to the cloud, where all the computing infrastructure and storage were available. This is the architecture shown in Figure~\ref{fig:arch-trad}. From a network point of view, there might be some gateway devices that take care of communication protocols (i.e., routing from the power-efficient local area network to the cloud); though from the logical point of view, they served no further purpose.

\begin{figure}
    \centering
    \includegraphics[scale=0.45]{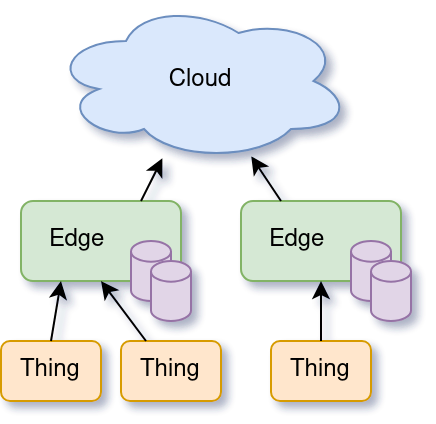}
    \caption{Continuum architecture with the presence of \emph{Edge} devices}
    \label{fig:arch-continuum}
\end{figure}

This original architecture, still valid for some scenarios, serves a clear purpose but presents some shortcomings: the volume data generated on the field can be substantially large, and the bandwidth and latency requirements of this architecture are challenging to support. To address these issues, the architecture evolved to include ``edge-devices'' and entertain more complex topologies \cite{ICOS-paper}. In general, nowadays, we can talk about \emph{computing continuum}\cite{balouek2019towards,abdelbaky2017computing}: a configuration that locates computing resources not only in the cloud but across the entire infrastructure. A stylized representation of this architecture can be seen in Figure~\ref{fig:arch-continuum}; take into account that nothing prevents a deeper hierarchy, some strategy that may be suitable for highly distributed and scaled environments\cite{ICOS-paper}.

\subsection{Leveraging the Active Storage System}

There are different strategies for defining the data management architecture within a computing continuum infrastructure. In this article, we will demonstrate the potential and benefits that \emph{active storage system} can bring by offloading artificial intelligence workloads across computing continuum resources.

We use the term \emph{active} storage systems to describe software systems that not only store data but also execute routines (functions associated with the stored data). An application using an \emph{active} storage system is able to persist data (into the storage system) and run methods (within the storage system, next to the data). In this kind of environment, we can identify two sides: client and server. The client is where all the application logic is executed, and the server (typically, a storage backend) is the location where the data is placed and where the code will be executed. In general, a single system can have multiple clients (multiple applications running) and multiple backends.

\begin{figure}
    \centering
    \includegraphics[width=0.4\textwidth]{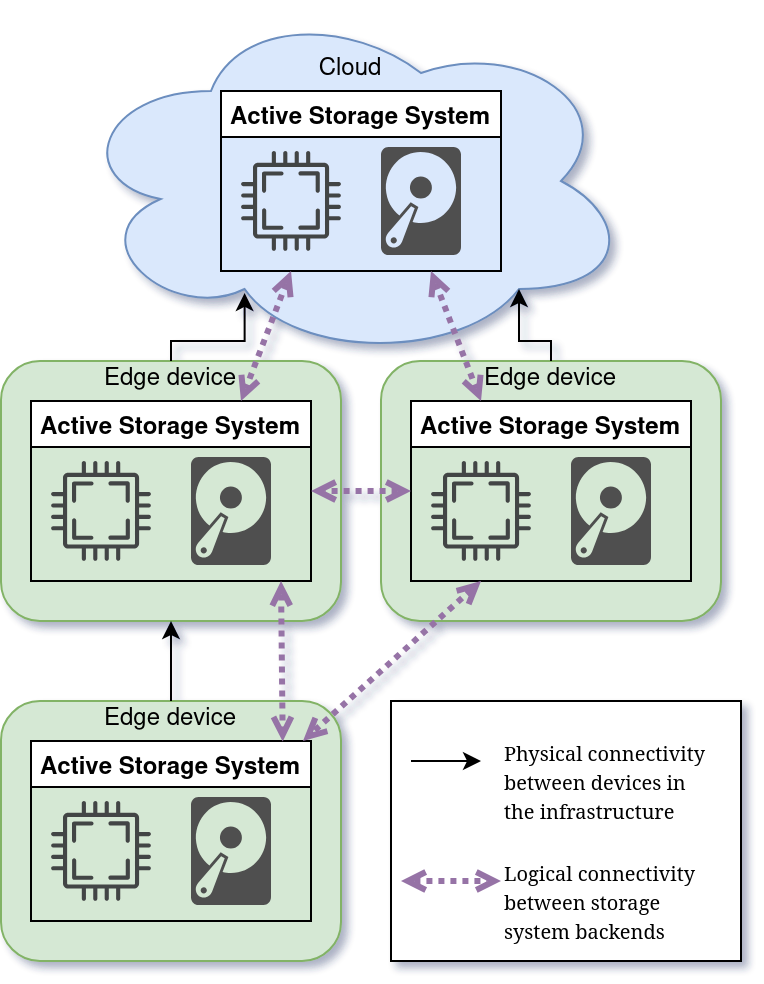}
    \caption{Continuum architecture including a distributed storage system across its edge and cloud nodes}
    \label{fig:arch-storagesystem}
\end{figure}

Figure~\ref{fig:arch-storagesystem} shows the architecture we propose. In this architecture, both edge devices and the cloud contain the active storage system component, a pivotal part of the data management that will manage the execution and storage of objects.

The novelty of this idea rests in the unified data management approach: a storage system capable of seamlessly running workloads in edge and cloud. This architecture can efficiently make use of continuum resources and perform offloading with low granularity.

\begin{figure}
    \centering
    \includegraphics[width=0.45\textwidth]{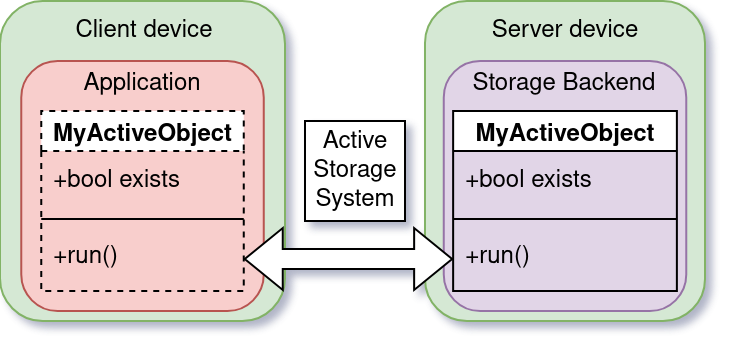}
    \caption{Detailed view on the logic relationship between the client and the server when considering objects of an active storage system. On the left, the application programming model interacts with an object that is a shadow of the real object, located and managed in the storage system backend (shown on the right).}
    \label{fig:arch-clientserver}
\end{figure}

A more in-depth example of how the programming model and the active object store interact is shown in Figure~\ref{fig:arch-clientserver}. In that figure, we can see an \textbf{Application}, running in what would be a client device. From the programming model point of view, the \textbf{Application} would interact with persistent objects (in the figure, a \texttt{MyActiveObject} instance, with its schema defined through the same programming model and including attributes and methods). The object available in the \textbf{Application} is a shadow of the object stored in the \textbf{Storage Backend}. This means that all interactions performed from the \textbf{Application} unto the \texttt{MyActiveObject} (on the left) are seamlessly being offloaded to the real instance in the \textbf{Storage Backend} (on the right). This procedure is performed by the active storage system and is transparent to the application and its programming model.

In this article, we will discuss and evaluate this architecture and the interactions of the programming model within the context of AI workloads. However, the proposed architecture could also benefit other domains with different workloads.

\subsection{Technical Implementation}
\label{technical-implementation}

The implementation that we propose will use dataClay \cite{dataclay-web}, a distributed active object store. This implementation intends to be a validation for our proposed architecture and will be used for evaluating its performance and behavior (Section~\ref{sec:results}). The chosen active object store is designed for both HPC environments and computing continuum infrastructures. Initially, it was born within the HPC ecosystem,  where its parallelism and scalability have been benchmarked up to several hundreds of cores, thousands of objects, and thousands of tasks on \cite{marti2017dataclay, barcelo2023enhancing}. In a continuum infrastructure, this translates to hundreds of edge devices running workloads simultaneously. In this context, dataClay has been deployed in real scenarios in different domains such as smart cities and smart transportation \cite{masip2021managing, serrano2021elastic, ICOS-paper} where it has proven to be a reliable and efficient data management solution. This distributed storage system can span several edge devices, managing the distribution of objects transparently to the application.

From a programmability point of view, dataClay presents a low entry barrier for application development. Objects in dataClay are presented through the native object-oriented interface and can be defined in the native programming language (Python). Methods of the object become part of the data model and can be executed transparently through the \emph{active} feature of the object store. This allows the domain expert to focus on the proper application logic and let the object store care about the distribution of objects and task offloading. Simultaneously, this system increases data locality, given that code execution (the methods) will be executed next to its data (the native programming language object and its attributes).

\begin{figure}
    \centering
    \includegraphics[width=0.48\textwidth]{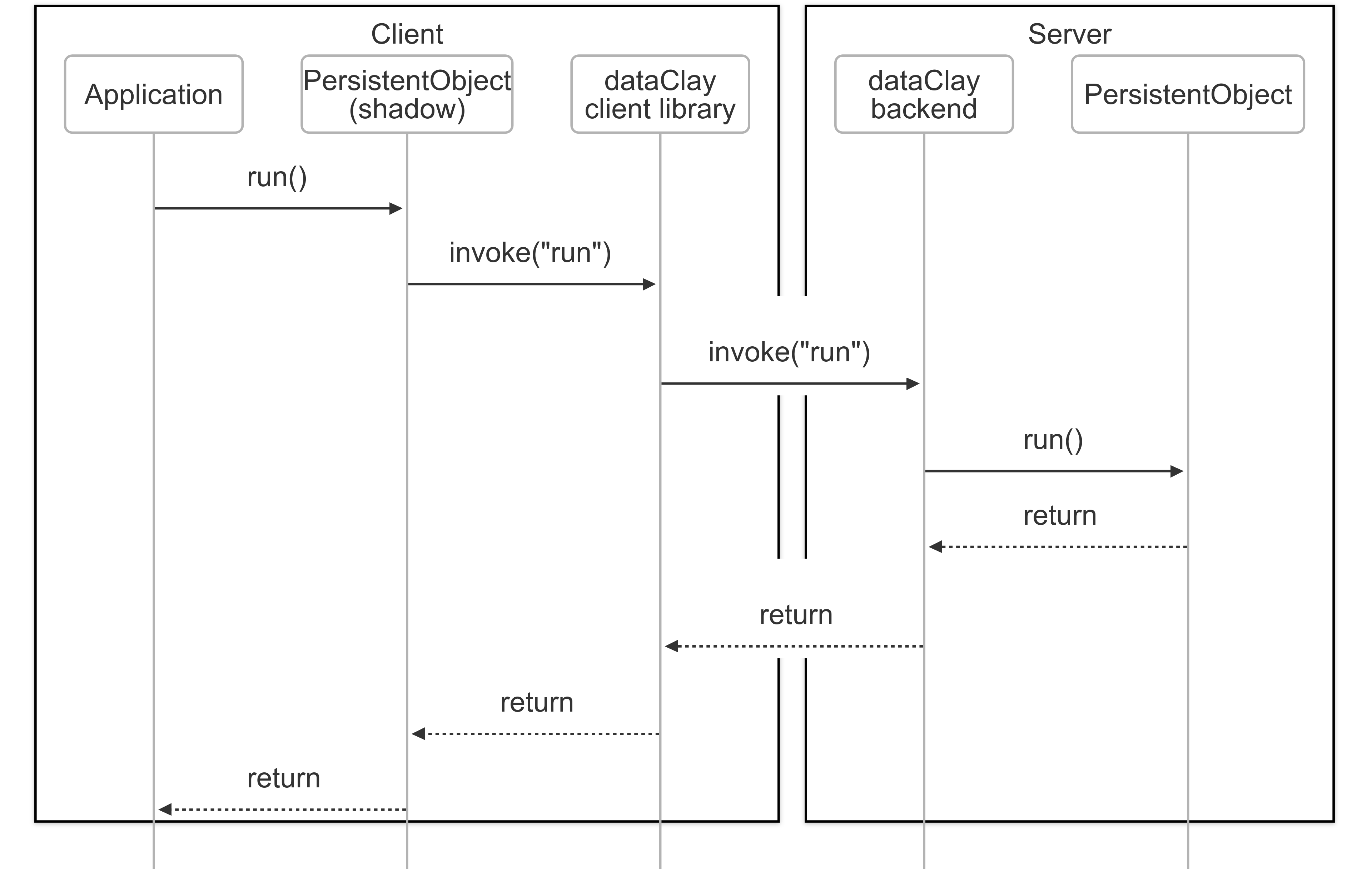}
    \caption{Sequence diagram of a \texttt{run()} invocation that starts in the Application (client side) and is being offloaded through dataClay into a Persistent Object in the Backend (server side).}
    \label{fig:sequencediagram}
\end{figure}

The high level behavior of the client-server interaction during an offloading procedure was previously shown in Figure~\ref{fig:arch-clientserver}. More specifically, we include here the sequence diagram of this process in Figure~\ref{fig:sequencediagram}. In this sequence diagram, we can observe how from the point of view of the \textbf{Application}, a \texttt{run} method is invoked and a return is received. The persistent object would leverage the dataClay client library and perform a remote call on the backend, where the \texttt{run} method is invoked in the actual object (the one that holds data).

This sequence diagram also demonstrates that dataClay architecture is compatible with real-time requirements. Given a task invocation on the client, a \texttt{run} in the client will result in an immediate \texttt{run} on the persistent object without the need of transferring data, and the return will be received as soon as the invocation finishes. Of course, there is a certain latency overhead due to the network communication, but the execution flow from the application is preserved without any rearrangement of tasks nor any asynchronicity.
Note, however, that it is virtually impossible to provide hard real-time guarantees in the continuum, as a result of the lack of strong network stability and the general need of plasticity across the cloud, edge and IoT. Thus, as commonly understood in continuum environments, the term \emph{real-time} here refers to soft real-time capabilities, which are mainly achieved by the fact that computation is performed where the data is, thanks to the active storage approach proposed.

When using dataClay, all application objects exist in a single scope and can be transparently accessed regardless of their location. This means that, from the point of view of the runtime and the developer, it does not matter if they are local objects or objects physically available in a remote backend: their object-oriented interface remains the same.

\begin{figure}
  \centering
  \begin{subfigure}[c]{0.48\linewidth}
    \includegraphics[width=\textwidth]{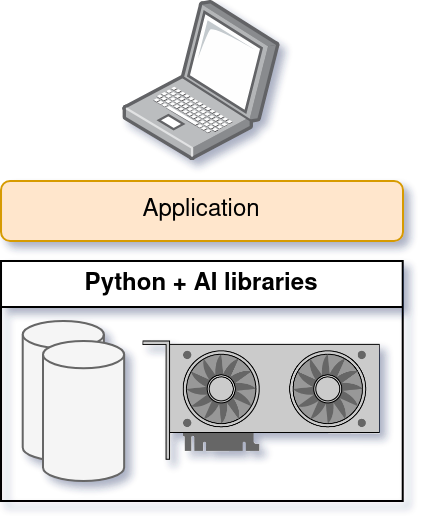}
    \caption{Scenario where the application is being run locally. This matches the scenario where a domain expert is developing an AI application and using local resources}
    \label{fig:app-arch_devel}
  \end{subfigure}
  \hfill
  \begin{subfigure}[c]{0.48\linewidth}
    \includegraphics[width=\textwidth]{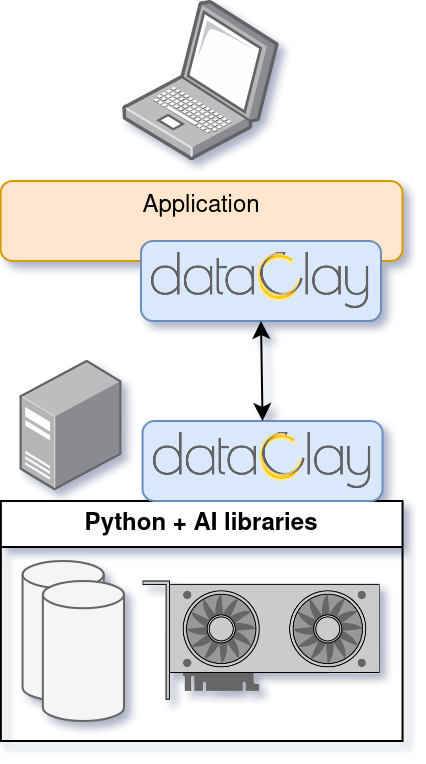}
    \caption{Scenario where the application is being run with dataClay. The application is the same, and we can see the connection between the dataClay client (top) and dataClay server (bottom). Resources being used are the ones on the server.}
    \label{fig:app-arch_dataclay}
  \end{subfigure}
  \caption{Two different scenarios where an application is using local resources (on the left) and where it is using remote resources (with dataClay, on the right)}
\end{figure}

Those dataClay characteristics are also invaluable for scaling applications from a single-machine development environment (Figure~\ref{fig:app-arch_devel}) to a distributed setting, where resources are remote (Figure~\ref{fig:app-arch_dataclay}). By abstracting the complexity of data distribution and remote execution, dataClay enables developers to write code as if all objects reside locally, reducing the cognitive overhead associated with managing data across multiple nodes. Additionally, its active storage capabilities ensure that computation is efficiently placed near the relevant data, minimizing communication overhead and improving performance. This makes it particularly advantageous for edge computing and cloud environments, where dynamic resource allocation and efficient data movement are critical for scalability and responsiveness.

If we look into the implementation details regarding the dataClay objects, we will see that an object is stored in the dataClay backend when it is created in it or when it is transferred to it. In dataClay terms, we say that an object is \emph{persistent} when it is physically in the memory of a backend. The application developer (or any higher controller layer used by the application) has control on where the objects are located. Objects can have replicas, and objects can be moved across the storage backends --creating replicas and moving objects are explicit user-space operations. In this article, we will consider deterministic placement of objects in order to focus the evaluation and reduce the undesired noise. However, the architecture proposed in this article is capable of accommodating all these dataClay mechanisms resulting in replicas and movements across a distributed continuum.

With dataClay as a storage system enabler, the transition from single-machine development to a distributed environment is seamless, as it preserves the object-oriented programming model without requiring modifications to existing libraries or coding paradigms. Developers can continue using the same programming language and tools while the system transparently handles data distribution and task offloading. This approach not only simplifies development but also ensures efficient resource utilization. These characteristics make dataClay a robust and flexible solution for our proposed architecture, facilitating deployment across edge and cloud infrastructures while maintaining high performance and programmability.

\subsubsection{\texttt{import} in constrained devices}
\label{stubobjects}

In an effort to be as transparent as possible, the data model of the application (i.e., the classes defined by the application) is used unmodified on the client side and the server side. Listing~\ref{lst:datamodel} shows the skeleton of a class, containing the required dataClay annotations and inheritance.

\clearpage

\begin{lstlisting}[language=python, style=modern, caption={Sample data model class definition for an AI application}, label=lst:datamodel]
from dataclay import DataClayObject, activemethod
import torch.nn as nn

class MyClass(DataClayObject):
  attribute: nn.Module

  @activemethod
  def method(...):
    pass
\end{lstlisting}

As explained in \ref{technical-implementation}, dataClay classes are defined through native Python classes (the \texttt{MyClass} definition in the example). Similarly, dataClay objects are represented through native Python objects. The schema definition (an \texttt{attribute} attribute that holds a PyTorch \texttt{Module} instance) is done through \texttt{typing} annotations in that same class. Finally, the active feature is defined through Python methods annotated with the \texttt{@activemethod} decorator.

From the \emph{client} point of view, an application using dataClay will connect to the dataClay services and leverage its features (persist objects, run active methods, etc.). A dataClay client library is responsible for performing all the remote calls under the hood while providing a clean and transparent interface to the developer.

In a dataClay deployment, the \emph{backend} is a service able to hold dataClay objects and run methods from those objects (i.e., run methods annotated with the \texttt{@activemethod}). A deployment can have one or more backends. By design, dataClay backends are implemented by running a Python interpreter, which holds objects in their native representation (as Python objects).

From a logical point of view, an application using the \texttt{MyClass} dataClay class will import the class (as would any Python application) and use it. During runtime, and once the object is persisted, that object will be transferred to the dataClay backend, and said backend will be a Python interpreter that will use that same class definition.

This way of developing applications and interacting with dataClay is very intuitive and practical for the application developer, avoids code duplication, and works well with code linters and other development tools. However, due to how Python \emph{import} works, this means that importing a class in the client will trigger all \texttt{import} lines defined in that file. This could potentially trigger the import of heavy libraries (e.g., the PyTorch library) that consume a lot of resources and have a long initialization time.

Note that there is no technical reason for requiring the chained import libraries --as long as only offloading operations and remote methods are used. For this reason, even while using dataClay, importing a class can result in heavy dependencies and unnecessary loading times. In an effort to provide a clean interface for applications that is suitable for constrained edge devices (devices where installing certain Python libraries is avoided), a feature called \emph{Stub objects} has been offered since dataClay 4.2.

When using dataClay stubs, the original class remains the same (i.e., the example Listing~\ref{lst:datamodel} does not change), but we change how to import dataClay classes from the client. The main application (on the client side) would use the \texttt{StubDataClayObject} as shown in Listing~\ref{lst:stubobject}.

\begin{lstlisting}[language=python, style=modern, caption={How an application can use a Stub object and avoid requirements on chained imports}, label=lst:stubobject]
from dataclay import DataClayObject, StubDataClayObject

# client-side, main application
MyClassStub = StubDataClayObject["mymodel.myclass"]
\end{lstlisting}

In this example, we see how \texttt{MyClassStub} can be used just as if it were the original \texttt{MyClass}, but it avoids an explicit import, so neither the data model nor its imports are required on the client side of the application.

This feature is used for the evaluation and has a direct and quantifiable impact on the resources used on the client side of the experiments.

\section{Methodology}
\label{sec:methodology}

We will perform a series of experiments in order to validate and evaluate the architecture proposed in Section~\ref{sec:arch} (Figure~\ref{fig:arch-storagesystem}). The following subsections will outline the specific AI workload to be tested and the experimental setup, including the hardware and software environments used for the evaluation.

\subsection{AI Workload} 

The workload that we will use for evaluation and validation is training an LSTM model (Long short-term memory, a type of recurrent neural network) with PyTorch. This is a straightforward intelligence application that can be used in many scenarios and may require scalability. These kinds of workloads can be common in the computing continuum, environments that may generate a lot of time-series data for a lot of metrics, devices, sensors, etc.

The models and procedures remained consistent throughout the experiments.

All the code necessary to replicate the experiments in this article is available on GitHub\footnote{
\url{https://github.com/sebasmos/EdgeAI-Continuum}
}.

\subsubsection{Dataset and Prediction Framework} 

\begin{figure}
    \centering
    \includegraphics[width=\linewidth]{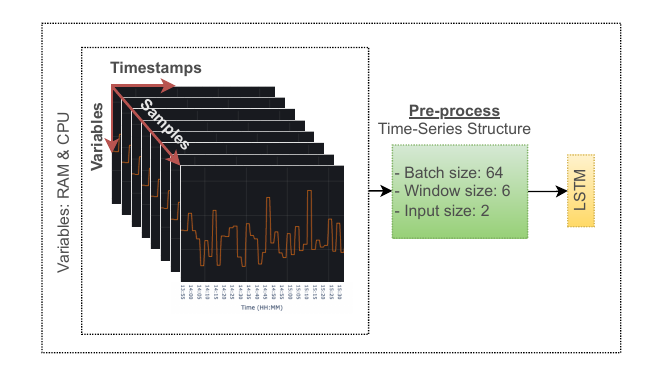}
    \caption{Data preprocessing pipeline showcasing the raw CPU and memory utilization data transformation into a structured multivariate time-series format for model training.}
    \label{fig:data-processing}
\end{figure}

The experiments were conducted on a dataset comprising two variables: historical CPU and memory utilization data collected from a Raspberry Pi 5 with 8GB RAM with a 64-bit quad-core Arm Cortex-A76 CPU, sampled every 5 minutes. As shown in Figure~\ref{fig:data-processing}, this scenario involves only two variables, but the workload can scale to accommodate more at higher dimensions.

The multivariate time-series data is structured into a supervised learning format using a batch size of 64, a window size of \( L = 6 \) lags in an autoregressive manner, and an input size of \( k = 2 \), representing the two covariates. At each time step \( t \), the model receives as input:

\[
X_t = \{x(t-1), x(t-2), \dots, x(t-L)\} \in \mathbb{R}^{L \times k}
\]

where \( x(t) = (x_1(t), x_2(t)) \) represents CPU and memory utilization. The model predicts the next step:

\[
Y_t = x(t)
\]

Since the data is sampled every 5 minutes, this corresponds to a 5-minute-ahead forecast. The data was normalized to the $[0,1]$ range and partitioned into training and validation sets (80\%-20\%). The dataset is publicly available on Hugging Face\footnote{\url{https://huggingface.co/datasets/ICOS-AI/synthetic_cpu_utilization}}.

\subsubsection{Model Architecture}
\begin{figure}
    \centering
    \includegraphics[width=\linewidth]{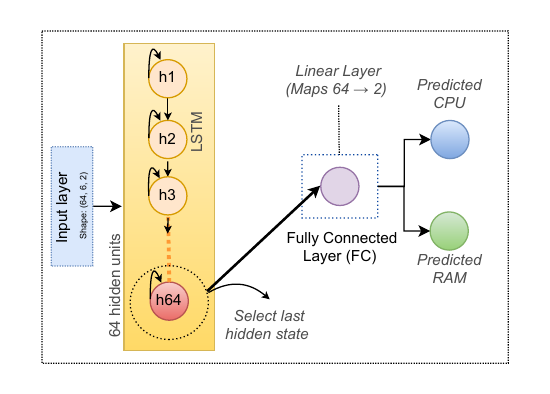}
    \caption{LSTM model architecture used for CPU and memory utilization prediction. The input layer processes sequences of shape $(64, 6, 2)$, which are passed through an LSTM layer with 64 hidden units. The final hidden state is selected and mapped to a 2-dimensional output via a fully connected layer.}
    \label{fig:model}
\end{figure}

The LSTM model captures temporal dependencies in the dataset covariates. As shown in Figure~\ref{fig:model}, the model consists of an LSTM layer with 64 hidden units, followed by a fully connected (FC) layer that maps the LSTM output to the final prediction. The model processes input sequences of shape $(64, 6, 2)$, where 64 represents the batch size, 6 is the look-back window, and 2 corresponds to the CPU and memory utilization covariates. 

The model was trained for 100 epochs using a batch size of 64, optimized with the Adam optimizer (learning rate = 0.001) and mean squared error (MSE) loss.

\subsection{Experimental Setup}

The evaluation in this article focuses on execution time and resource usage (system memory and disk requirements), and these metrics will be evaluated in all the different experimental setups. These experimental setups can be divided into two sets:

\begin{description}
    \item[Baseline] experiments with no dataClay, run locally, to have a measure of the maximum performance given a certain hardware.
    \item[dataClay] experiments including a client-server approach in which a \emph{client} device is using the dataClay active mechanism to offload the AI workload to a \emph{server} device (which is running a dataClay backend).
\end{description}

We will use three different hardware devices for the evaluation: \textbf{Ryzen}, \textbf{Mac}, and \textbf{OrangePi}. The characteristics of those devices are detailed in Section~\ref{sec:hardware}.

There will be two \textbf{Baseline} experiments, one for each hardware edge device: \textbf{OrangePi} and \textbf{Mac}.

The \textbf{dataClay} experiments will cover all pairs (enumerated in the form \emph{server}-\emph{client}):
\begin{itemize}
    \item Ryzen-Mac
    \item Ryzen-OrangePi
    \item Mac-OrangePi
\end{itemize}

These experiments cover the different configurations to illustrate the behavior of the offloading mechanisms, from the point of view of the AI workload, depending on the resources available in the edge. In all dataClay experiments object placement is statically decided, so that objects are placed in the backend to be analyzed.

In particular, for all dataClay experiments, we will observe memory consumption and execution times. These metrics will be broken down into three different points of view:

\begin{enumerate}
    \item \textbf{dC-Client}: Point of view of the client triggering the application workload. From this point of view, we will discuss the memory used in the client device, the time consumed in the client library (including time consumption due to serialization, offloading, transfer, and other overheads).

    \item \textbf{dC-Server}: Point of view of the server. From this point of view, we will discuss the memory used by the dataClay backend (in the server device) and the time consumed during the offloaded operation.

    \item \textbf{dC-Agg}: The global point of view for both the client and server. The memory consumption is the sum of the two previous ones (giving the aggregated use of memory in the system). The execution time is the total execution time from when the client starts the workload until it receives the result (all serialization, transfer, and overheads as well as the processing time at the server are included in these measurements).
\end{enumerate}

Note that the third point of view is technically redundant, but including it in the tables and the evaluation will help contextualize the results and ease certain comparisons.

All experiments are run using 20 different random seeds to ensure robustness and reliability in the results. The PyTorch seeding mechanism was utilized to maintain consistency across these experiments, allowing for an accurate comparison of performance metrics while mitigating the effects of randomness inherent in training processes.

For all experiments, including dataClay, we will make use of its active features (for offloading the methods to the server). We will also use the data model through the Stub Object feature; this allows us to avoid installing all the big and resource-hungry specialized intelligence libraries, namely \texttt{torch}, \texttt{torchvision}, and \texttt{scikit-learn}. The insights and discussion will focus on both \textbf{Client} side (i.e., the part of the application triggering the training process) as well as the \textbf{Server} side (i.e., the methods of the data model, the part of the application performing the training itself). Note that this difference is relevant for evaluating the resource usage; however, this will be transparent for the domain expert- the one programming the application. The interface required to use dataClay is the object-oriented interface, the one already being used in the rest of the application. Thanks to this approach, there is no need to learn a new programming model, nor is it required to know or care about the exact topology of the available resources.

\subsubsection{Hardware Specifications}
\label{sec:hardware}

As mentioned before, three different hardware devices will be used for the evaluation. These three devices represent a wide range of performance, cost, and size characteristics. From less powerful to more powerful, those are the hardware devices that we will be using:

\begin{description}
    \item[OrangePi] The Orange Pi 5 single-board computer. Powered by a RK3588S 8-core processor (quad-core A76 and quad-core A55)
    \item[Mac] Equipped with 16 GB RAM and the Apple M2 Pro chip, the Mac boasts a 12-core CPU.
    \item[Ryzen] 32-core AMD Ryzen Threadripper PRO 5975WX. 
\end{description}

In a computing continuum context, the first two devices (\textbf{OrangePi}, a small and power-efficient device, and \textbf{Mac}, standing as a mid-sized device) can be considered possible edge devices. The \textbf{Ryzen} is a powerful hardware, which represents a cloud endpoint or a very powerful edge device.

Experiments have been run to simulate an edge-to-cloud IoT architecture. Edge devices, located near data sources, handle initial data collection. Due to their limited processing and storage, task offloading to the cloud may be desirable \cite{ICOS-paper}. 
dataClay facilitates the seamless transfer of these demanding workloads across the network. 
Thus, our goal is to demonstrate the ability to effectively offload computationally intensive tasks from resource-constrained edge devices. The evaluation likely focuses on proving the value of dataClay in simplifying and optimizing workload management in this distributed system.

The experiments will not exhaust the available resources, but evaluating these widely different devices is useful to gain insights into the growth capabilities and scalability limits of this kind of computing continuum architecture.

In a computing continuum infrastructure, many different factors have to be considered simultaneously (network infrastructure, computing resources, and workload variability, among others). Some of these variables are external (like workload variability) or difficult to modify (i.e., changing the network technology can result in a chain reaction where lots of hardware devices must be changed to ensure compatibility). In contrast, the flexibility of continuum infrastructures allows adding or substituting any edge device, which may have a clear impact on execution time. Thus, we will focus on the evaluation of  the total execution time of an AI workload, with and without task offloading, for several computing resource configurations.

\subsubsection{Software and Libraries}

The active storage system that we will use is dataClay\cite{dataclay-web} version 4.2 (beta). The 4.2 version includes the \emph{Stub Object} feature described in Section~\ref{stubobjects}, a feature that is relevant for the client-side measurements in dataClay experiments.

The described AI workload is a Python application that will be run on Python 3.10. The requirements of this application include:

\begin{itemize}
    \item A set of basic scientific libraries, including \texttt{numpy} (version 1.24.2) and \texttt{pandas} (version 1.5.3)
    \item Specialized intelligence libraries: \texttt{torch} (version 2.0.0), \texttt{torchvision} (version 0.15.1) and \texttt{scikit-learn} (version 1.2.2)
    \item Some additional libraries for configuration and performance monitoring (e.g., \texttt{memory-profiler}, \texttt{seaborn} and \texttt{notebooks})
\end{itemize}

\section{Results and Analysis}
\label{sec:results}

In this section, we will discuss the results obtained from our experiments, following the methodology explained in the previous Section~\ref{sec:methodology}. These experiments show the behavior of different hardware resources in the context of continuum infrastructures and their trade-offs.

The following tables show the memory and runtime performance of the different experiments. Table~\ref{tab:memory_runtime_baseline} shows the memory consumption and execution times of the two baseline experiments (execution without dataClay). There is an experiment for each edge device: Mac and OrangePi.

\begin{table}[htbp]
    \centering
    \scriptsize 
    \setlength{\tabcolsep}{2pt}
    \begin{tabular}{p{2cm} p{2cm} p{1.5cm} p{1.5cm} p{1cm}}
        \toprule
        \textbf{Hardware} & \textbf{Memory (MB)} & \textbf{Train Time} & \textbf{Eval Time} & \textbf{Total Time} \\
        \midrule
        
        Mac
          & $398.13 \pm 2.19$ 
          & $6.18 \pm 0.11$ 
          & $0.84 \pm 0.03$ 
          & $7.02$ \\

        OrangePi
          & $367.12 \pm 1.9$
          & $37.2 \pm 0.4$
          & $4.32 \pm 0.23$
          & $41.52$ \\
          
        \bottomrule
    \end{tabular}
    \caption{\textbf{Baseline} -- Memory and runtime performance for the Baseline experiments (run on edge devices). Memory values are reported with their standard deviation using ± notation.}
    \label{tab:memory_runtime_baseline}
\end{table}

Looking into the values shown in Table~\ref{tab:memory_runtime_baseline}, we can observe that memory usage in both hardware is comparable (the Mac shows a slightly higher memory consumption). Unsurprisingly, the more powerful device (Mac) completes the overall process significantly faster (7.02~s compared to the 41.52~s on the OrangePi), highlighting a substantial performance advantage. More specifically, the \emph{training} portion on the Mac is about six times faster; the training represents most execution time relative to the total.

It is important to mention how both dataset size and model complexity will simultaneously affect execution time and memory usage. The device characteristics (the hardware where the workload is run) will mostly affect the execution time. However, in order to provide a meaningful comparison, all the experiments maintain the dataset and model constant.

The three dataClay experiments, which perform offload of intelligence workloads across devices, are shown in individual tables:

\begin{itemize}
    \item Table~\ref{tab:memory_runtime_dcryzenmac}: Ryzen-Mac configuration.
    \item Table~\ref{tab:memory_runtime_dcryzenorangepi}: Ryzen-OrangePi configuration.
    \item Table~\ref{tab:memory_runtime_dcmacorangepi}: Mac-OrangePi configuration.
\end{itemize}

\begin{table}[htbp]
    \centering
    \scriptsize 
    \setlength{\tabcolsep}{2pt}
    \begin{tabular}{p{2cm} p{2cm} p{1.5cm} p{1.5cm} p{1cm}}
        \toprule
        \textbf{Scope} & \textbf{Memory (MB)} & \textbf{Train Time} & \textbf{Eval Time} & \textbf{Total Time} \\
        \midrule
        
        dC-Client
          & $304.78 \pm 0.94$ 
          & -- 
          & -- 
          & 6.47 \\
          
        dC-Server
          & $490.44 \pm 5.41$ 
          & $4.11 \pm 0.11$ 
          & $0.45 \pm 0.09$ 
          & $4.57$ \\
          
        dC-Agg
          & $795.22 \pm 5.49$ 
          & -- 
          & -- 
          & $11.04$ \\
          
        \bottomrule
    \end{tabular}
    \caption{\textbf{dataClay Ryzen-Mac} experiment -- Memory and runtime performance across the dataClay experimental setup between Ryzen (server) and Mac client. Memory values are reported with their standard deviation using $\pm$ notation.}
    \label{tab:memory_runtime_dcryzenmac}
\end{table}

\begin{table}[htbp]
    \centering
    \scriptsize 
    \setlength{\tabcolsep}{2pt}
    \begin{tabular}{p{2cm} p{2cm} p{1.5cm} p{1.5cm} p{1cm}}
        \toprule
        \textbf{Scope} & \textbf{Memory (MB)} & \textbf{Train Time} & \textbf{Eval Time} & \textbf{Total Time} \\
        \midrule
          
        dC-Client
          & $329.55 \pm 0.02$ 
          & -- 
          & -- 
          & $8.33$ \\
          
        dC-Server
          & $480.18 \pm 4.70$ 
          & $4.12 \pm 0.03$ 
          & $0.45 \pm 0.07$ 
          & $4.58$ \\
          
        dC-Agg
          & $809.73 \pm 4.70$ 
          & -- 
          & -- 
          & $12.91$ \\
          
        \bottomrule
    \end{tabular}
    \caption{\textbf{dataClay Ryzen-OrangePi} experiment -- Memory and runtime performance across the dataClay experimental setup between Ryzen (server) and OrangePi (client). Memory values are reported with their standard deviation using $\pm$ notation.}
    \label{tab:memory_runtime_dcryzenorangepi}
\end{table}

\begin{table}[htbp]
    \centering
    \scriptsize 
    \setlength{\tabcolsep}{2pt}
    \begin{tabular}{p{2cm} p{2cm} p{1.5cm} p{1.5cm} p{1cm}}
        \toprule
        \textbf{Scope} & \textbf{Memory (MB)} & \textbf{Train Time} & \textbf{Eval Time} & \textbf{Total Time} \\
        \midrule  

        dC-Client 
          & $322.57 \pm 0.04$ 
          & -- 
          & -- 
          & $15.98$ \\
          
        dC-Server
          & $389.77 \pm 2.07$ 
          & $15.27 \pm 1.78$ 
          & $1.26 \pm 0.22$ 
          & $16.53$ \\
          
        dC-Agg
          & $712.35 \pm 2.07$ 
          & -- 
          & -- 
          & $32.50$ \\
          
        \bottomrule
    \end{tabular}
    \caption{\textbf{dataClay Mac-OrangePi} experiment -- Memory and runtime performance across dataClay experimental setup between Mac (server) and OrangePi (client). Memory values are reported with their standard deviation using $\pm$ notation.}
    \label{tab:memory_runtime_dcmacorangepi}
\end{table}

Those tables summarize memory and runtime performance across dataClay configurations. Experiments using a Ryzen server (Tables~\ref{tab:memory_runtime_dcryzenmac},\ref{tab:memory_runtime_dcryzenorangepi}) achieved significantly lower training (4.1s) and evaluation (0.45s) times than those using a Mac server (15.3s and 1.26s respectively, Table~\ref{tab:memory_runtime_dcmacorangepi}).

Table~\ref{tab:key_metrics} shows the performance of the trained LSTM model. The table shows the following key performance metrics:

\begin{description}
\item[MSE] Mean Squared Error
\item[MAE] Mean Absolute Error
\item[SMAPE] Symmetric Mean Absolute Percentage Error
\item[RMSE] Root Mean Squared Error
\end{description}

\begin{table}
    \centering
    \scriptsize 
    \setlength{\tabcolsep}{4pt} 
    \begin{tabular}{l c c c c}
        \toprule
        \textbf{Variable} & \textbf{MSE} & \textbf{MAE} & \textbf{SMAPE} & \textbf{RMSE} \\
        \midrule
        $x_1(t)$ [CPU] & $7.651 \pm 0.539$ & $2.115 \pm 0.107$ & $6.168 \pm 0.329$ & $2.764 \pm 0.097$ \\
        $x_2(t)$ [Mem] & $0.871 \pm 0.353$ & $0.733 \pm 0.197$ & $0.835 \pm 0.224$ & $0.914 \pm 0.192$ \\
        \bottomrule
    \end{tabular}
    \caption{Key Performance Metrics Across Experiments. MSE, MAE, SMAPE, and RMSE are shown as representative metrics, with Model Size omitted as it remains constant at \textbf{0.0707 MB}.}
    \label{tab:key_metrics}
\end{table}

Ideally, accuracy should remain consistent across all setups, with minor variances due to different random seeds used in computations. The evaluation revealed global averages for key metrics: The memory-based setup consistently yields lower error values (MSE, MAE, SMAPE, and RMSE) compared to the CPU configuration, indicating superior predictive performance when using memory while maintaining a model size of \textbf{0.0707 MB}. Although the CPU shows higher absolute errors, its lower relative standard deviations indicate higher stability, while the memory configuration exhibits higher variability.

The following subsections will discuss each evaluation metric: memory usage (\ref{subsec:memory_usage}), execution time (\ref{subsec:execution_time}), and storage requirements (\ref{subsec:storage_req}).

\subsection{Memory Usage}
\label{subsec:memory_usage}

\begin{figure}
    \centering
    \includegraphics[width=\linewidth]{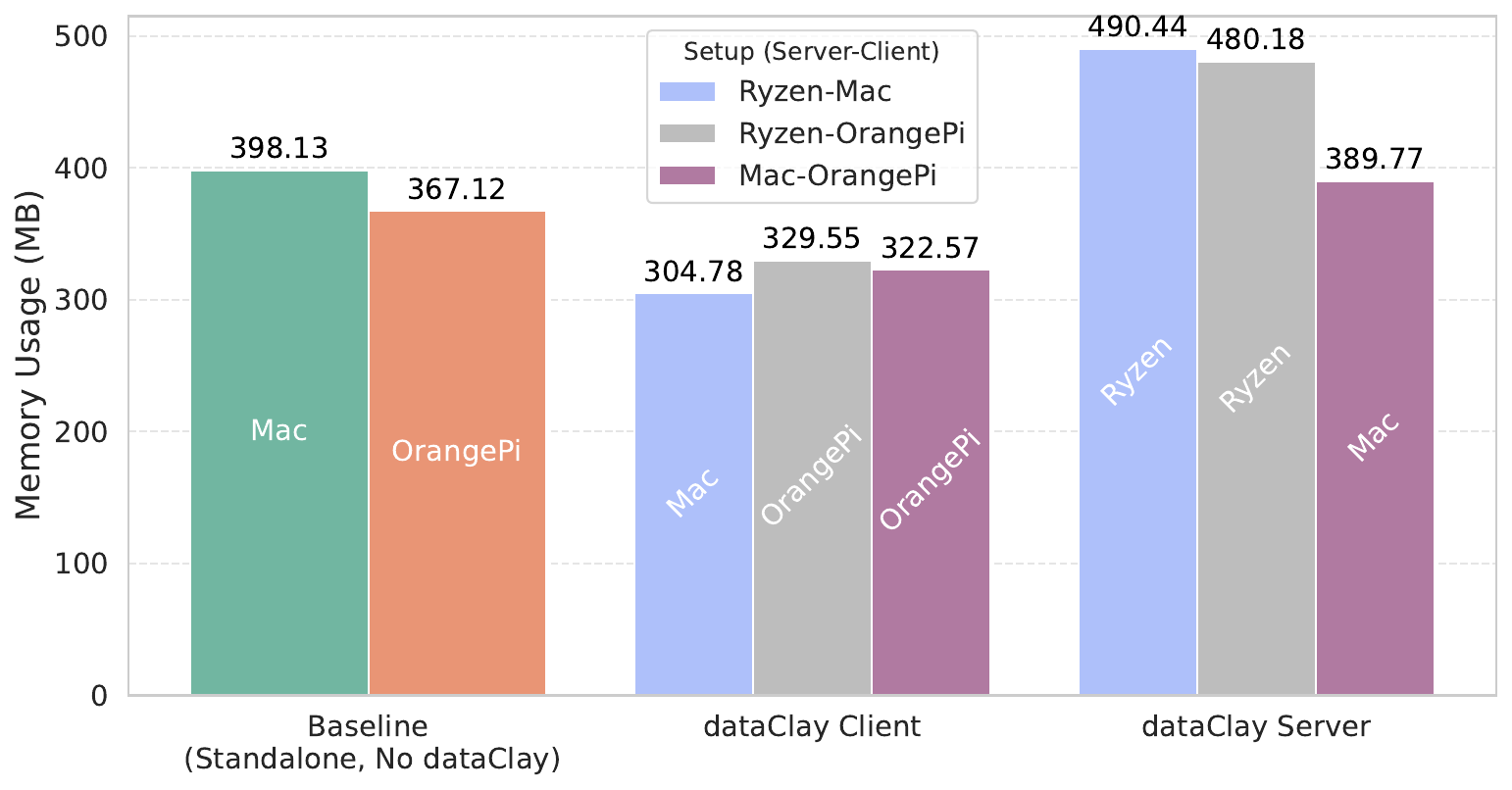 }
    \caption{Memory Comparison: Baseline vs dataClay}
    \label{fig:mem_plot}
\end{figure}

Figure~\ref{fig:mem_plot} illustrates memory usage, showing first the baseline (with no dataClay, for Mac and OrangePi devices), followed by the memory usage on the client and on the server device in all dataClay experiments.

If we examine these results from the perspective of the continuum infrastructure, we can observe a clear reduction in memory requirements on the client compared to the baseline. Because that would be the smaller edge device, where we may typically observe the deployment of constrained devices, this is a desirable result. It shows the flexibility of the active storage system in a computing continuum: we can reduce the hardware requirements on constrained edge devices.

These experiments need a server counterpart that will also consume resources. However, the resources consumed are comparable, as we can see that the Mac server has the same memory consumption as the Mac standalone (something that we were already expecting, given that they both are effectively performing the same workload). The higher core count on the Ryzen results in a higher memory consumption, but it also comes with decreased execution times, as we will later discuss. Ryzen is shown here as an example of cloud infrastructure where resources are expected to be more abundant (although it could also be understood as a very powerful edge device).

\subsection{Execution Time}
\label{subsec:execution_time}

\begin{figure}
    \centering
    \includegraphics[width=\linewidth]{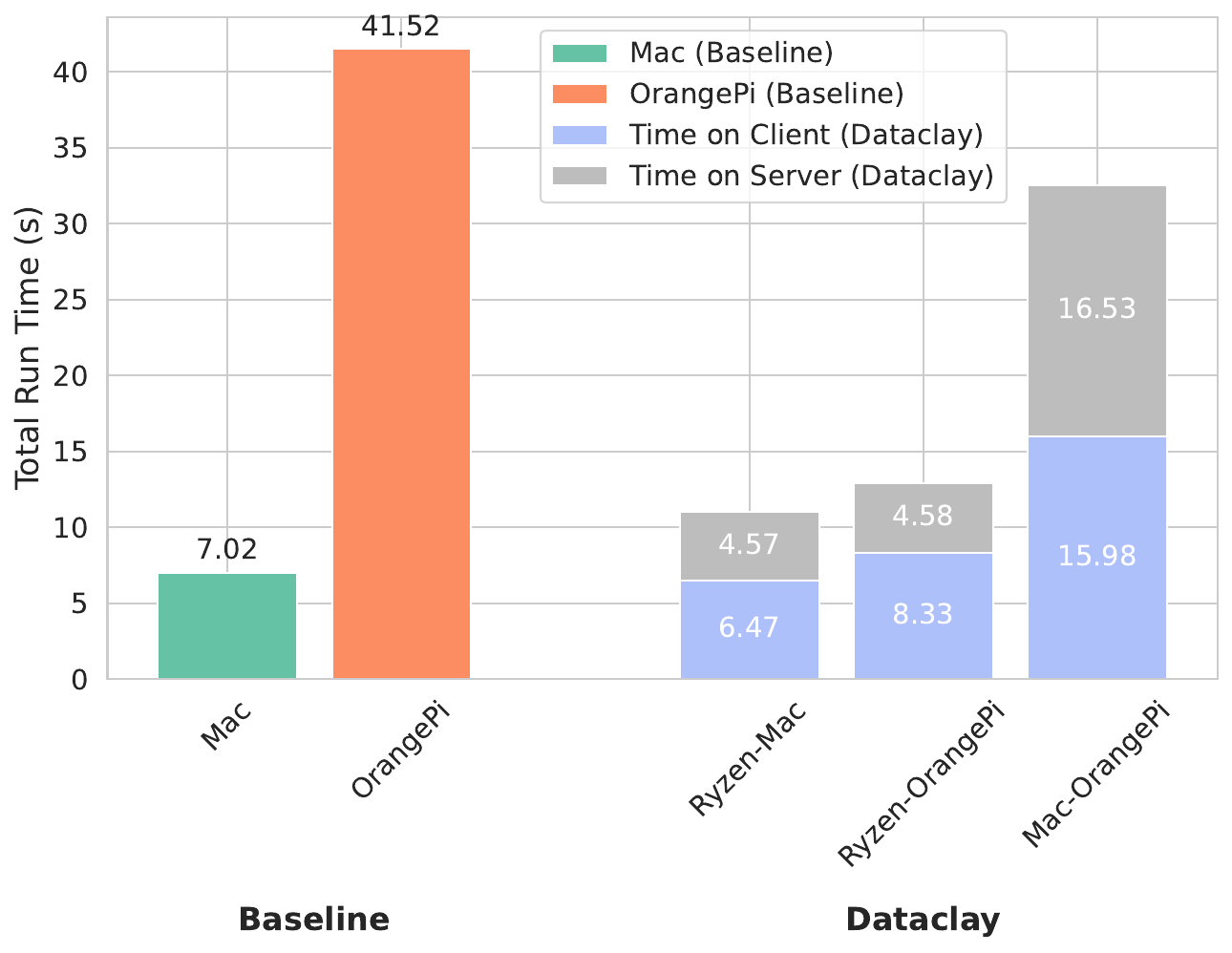 }
    \caption{Run Time Comparison: Baseline vs dataClay}
    \label{fig:runtime_plot}
\end{figure}

Figure~\ref{fig:runtime_plot} shows execution times. There is a general trend related to the power of the device: bigger hardware is faster. This holds true both for the baseline and for dataClay experiments.

The baseline executions are a straightforward measurement of the application time. The dataClay bars are itemized into server time (execution time of the task itself) and client time (all the task offloading overhead, including dataClay library execution, serialization, deserialization, transfer, etc.).

Seeing how more powerful devices run the same workload faster is not a surprising result. It is interesting to corroborate that, among dataClay experiments, 
a key factor in the execution time is the server device running the workload --while the client hardware has a lower impact. This is a result that we expected given that the model is being trained and evaluated on the server; the training and evaluation time is an important portion of the total time (as shown in the tables in Section~\ref{sec:results}).

Having a client-server architecture does have a certain overhead, and that overhead is indeed affected by the client device (this is the reason behind Ryzen-Mac being faster than the Ryzen-OrangePi, even when the training and evaluation are performed in the same Ryzen). We have to take into account the serialization cost, network, and remote calls; i.e., all the overhead of distributed data management. All that client-server overhead (which is the main contributor in ``Time on Client'' for the dataClay configurations) will greatly depend on hardware characteristics (network connectivity, computing resources). In these experiments we show a controlled scenario with direct connectivity, and we focus the evaluation on the different computing device behaviors (a reasonable degree of freedom on continuum infrastructures) and we maintain a constant network. Very constrained networks or high-latency environments would inevitably result in higher ``Time on Client'' measurements.

If speed is paramount, the best solution is to run the workload locally on a powerful device. However, that is not an economic solution and does not scale in computing continuum environments. Using small, affordable, power-efficient devices (such as the OrangePi) as clients to a more powerful device is a cost-effective infrastructure design. Active storage systems present a flexible way to exploit resources in the infrastructure in a transparent way, while maintaining programmability.

\subsection{Storage Requirements}

\label{subsec:storage_req}

An aspect that is often overlooked when using frameworks is their storage requirements, i.e., how big are the libraries/containers involved in the software architecture. In a lot of distributed environments, the framework size would be small compared to the workloads and other storage requirements of the system. However, in computing continuum environments, edge devices can be small devices with small and low-endurance flash memories, and the software can be a relevant part of the total available storage capacity of those devices.

Table~\ref{tab:storage_usage} shows the storage requirements of the software architecture. For the baseline, we see that about 2GB are required by the application and its ML libraries. This is a sizable amount of the storage available in, say, an OrangePi device with a limited flash drive attached to it. Especially so if we consider that the edge device may have other frameworks and tools required by its computing continuum purpose.

\begin{table}[htbp]
    \centering
    \begin{tabular}{l c}
        \toprule
        \textbf{Setting} & \textbf{Size (GB)} \\
        \midrule
        Baseline & 2.1048 \\
        dataClay (client) & 0.4283 \\
        dataClay (server) & 7.0048 \\
        
        \bottomrule
    \end{tabular}
    \caption{Storage Comparison Across Configurations}
    \label{tab:storage_usage}
\end{table}

By using an active storage system, we eliminate the need for heavy ML libraries on the client, allowing us to dramatically reduce storage requirements on the client-side devices of the infrastructure. For dataClay, the table shows that less than 0.5GB is required for the application.

The client application will not have those ML libraries, but the dataClay goal is to offer a transparent programming interface with a low entry barrier. By virtue of the feature explained in Section~\ref {stubobjects}, the Python objects will behave as if they were local, but the object will remain on the server. Execution will be performed next to its data, and the client does not need to have the libraries installed.

Of course, we need to account for the storage requirements on the server. The active storage system that we are using, dataClay, has a certain number of dependencies and extra requirements (on top of the ML libraries seen in the baseline), thus the storage required by the server increases to 7GB. However, these 7GB include all the additional features of the storage system (for instance, the capacity to store persistent data and share datasets between clients). 

The storage comparison we provide is not under identical conditions, and it is unfair for dataClay: the measurements done result in a higher number of dependencies for the dataClay (server) configuration (resulting in a higher size value). Our goal was to compare general storage requirements and variability, and that was done by analyzing storage in the \texttt{site-packages} folder (the place where Python libraries are installed). However, the baseline is installed in the host with Conda (environment manager, on bare metal) while the dataClay (server) configuration considers the usage within the containerized image, where packages are installed through vanilla \texttt{pip install}. The PyTorch framework (\texttt{torch} package) installs differently in certain environments and certain installation options. In a regular development desktop, we can see how a \texttt{pip install torch} command results in a \texttt{site-packages} of 5.1~GB, which is much greater than the baseline. This discrepancy is due to the fact that the default installation of the \texttt{torch} package from PyPI installs CUDA requirements and many extras. It is perfectly feasible to build a lighter server dataClay image (trimming those extra requirements), but that would also result in a less versatile dataClay container. The infrastructure operator may decide on having multiple lighter images and deploy them specifically, or rather have a single unified (``conservative'') image with all the requirements --paying the price of the increase in storage requirements. The comparison in this article shows the conservative approach, the worst-case scenario.

Reducing the requirements for the most constrained devices is a positive result. Independently of the storage required on the server side, we have shown how an active storage system can achieve those results without changing the programming paradigm of the application. The increase of storage requirements (in the server side) is an effect that is diluted (``shared'') among all the clients in the infrastructure, so we can affirm that using the proposed architecture quickly results in a net reduction in overall storage for multi-client infrastructures.

\section{Towards Distributed Workloads}
\label{sec:towards-distributed}

The experiments and evaluation in this article have, so far, focused on device heterogeneity and point-to-point offloading. This scenario is a common one in AI workloads across the continuum and demonstrates the key strengths of our proposed architecture. Moreover, it perfectly fits the behavior of \texttt{torch}, a widely used framework and also the one used in our demonstration application (specifically, within the LSTM model).

However, it is also relevant to analyze the capabilities and the potential of active storage systems for highly scalable distributed machine learning applications. Thus, we provide an additional set of experiments evaluating a naturally distributable workload that can be run in an homogeneous HPC infrastructure, which allows us to focus on the distribution aspect and the scalability of the software stack.

\subsection{Software Infrastructure and Application}

The algorithm selected for the experiments is the Cascade SVM\cite{graf2004parallel} (or CSVM), an algorithm for Support Vector Machines that can be parallelized efficiently and scales to very large problems. The algorithm itself is designed to accommodate the parallelism and computation distribution. In particular, we will be using the \texttt{dislib}\cite{dislib} implementation of this algorithm. The \texttt{dislib} library contains several distributed algorithms in Python, ready to use. It is highly focused on machine learning algorithms (and greatly inspired by scikit-learn). The support for task distribution is provided by the COMPSs task-based programming model\cite{tejedor2017pycompss}, a distributed framework that can be used in HPC, cloud and continuum environments. More specifically, we will be using PyCOMPSs, its Python bindings.

We will compare the execution performance and scalability with and without dataClay in order to assess the impact that the active storage system brings to a distributed workload such as the one described in this section.

\subsection{Hardware Infrastructure}

The experiments will be performed in an HPC environment, which allows us to focus on the distribution aspect of the execution. In general, HPC environments will have high bandwidths and robust network infrastructures, which is perfect for isolating the distribution aspects from jitter and noise from unreliable network connectivity.

All the experiments are executed in the MareNostrum 4 HPC cluster~\cite{marenostrum}. The nodes in this cluster have the following technical specs:

\begin{itemize}
    \item 2$\times$Intel\textregistered{} Xeon\textregistered{} Platinum 8160L CPU @~2.10GHz
    \item 96GB of DRAM (12$\times$8GB 2667MHz DIMM)
    \item 100 Gb/s Intel Omni-Path (between computing nodes)
    \item 10 Gb Ethernet (storage and management)
\end{itemize}

Each node contains a total of 48 ($2 \times 24$) cores. Given that each node is very powerful and there is certain independence between the processors (memory lanes, internal caches, etc.), we will focus on the number of processors. The active storage system (in those experiments where it is present) will be deployed with a backend on each processor, with \texttt{numactl} ``isolation'' from its neighbor processor.

The experiments will be done from 1 to 16 nodes, or equivalently, from 2 to 32 processors. This results in a maximum of 768 cores (32 processors$\times$24 cores per processor). The total number of cores is the concurrent number of tasks that COMPSs will schedule. This results in a good indicator of distributed performance and behavior of our software stack.

\subsection{Dataset and Methodology}

The experiments showcase the weak scaling performance of the Cascade SVM for a synthetic dataset of 150 thousand points per processor. The evaluation will be performed by a weak scaling up to 32 processors (16 nodes).

The first set of experiments represent a highly fragmented dataset. The execution is performed with a block size of 128 points. This results in a workload of 8 blocks per core, or 192 blocks per processor.

The second set of experiments increases the block size in order to reduce the stress on the scheduler and balance the number of blocks to the number of cores, maintaining the total dataset size. This change reduces the compute overheads due to scheduling and reduction operations, and it allows us to focus on the data management costs in the software stack.  The block size is increased from from 128 to 1024 which results in a workload of 1 block per core, or 24 blocks per processor.

\subsection{Results and Analysis}

\begin{figure}
    \centering
    \includegraphics[width=0.95\columnwidth]{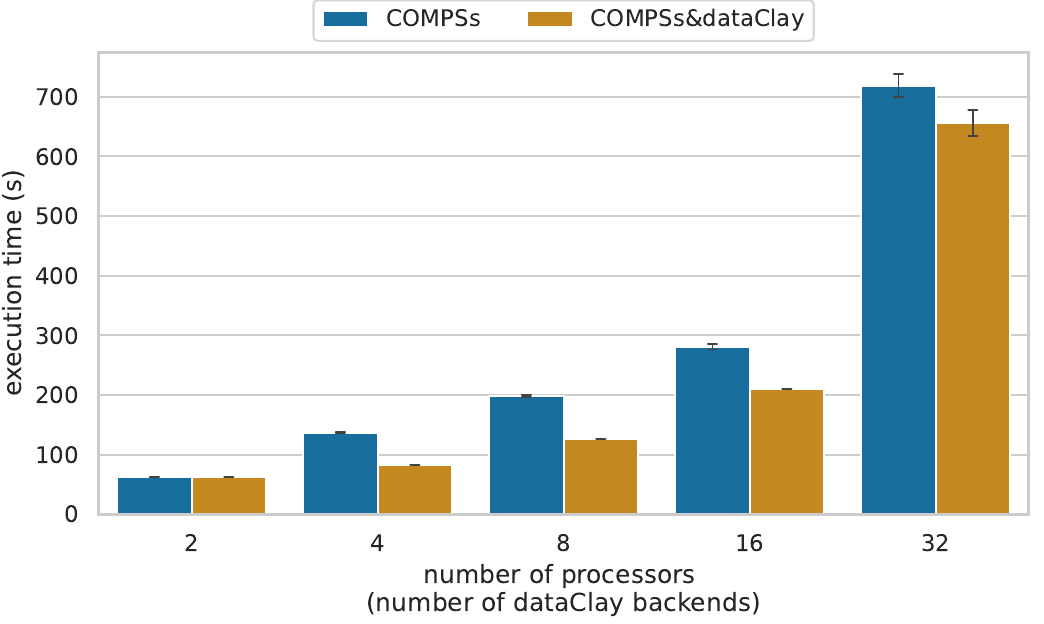}
    \caption{Cascade SVM weak scaling from 2 processors (1 node, 2 backends) to 32 processors (16 nodes, 32 backends), with 192 blocks per processor (8 per core)}
    \label{fig:csvm-weakscaling_smallblocks}
\end{figure}

The weak scaling experiment for the highly fragmented dataset (192 blocks per processor) is shown in Figure~\ref{fig:csvm-weakscaling_smallblocks}. The general scalability of the system is consistent up to 16 processors, but we can observe that performance for 32 processors is clearly degrading when compared to smaller experiments. Given that this is a highly fragmented dataset, with a lot of small blocks and a lot of short distributed tasks, it is expected to see some eventual degradation (the weak scaling scales the dataset, which scales in the embarrassingly parallel stages of the algorithm, but the aggregation complexity increases with the size of the problem). Given the sheer number of tasks and blocks, we can conclude that there is an overall good behavior and trend due to the COMPSs framework but there is apparent degradation starting at 32 processors.

Having dataClay in the software stack results in a significant performance improvement. When dataClay is available, task execution exploits data locality and the distributed execution is leveraging the active aspect of dataClay. This is able to reduce certain data movements and thus improves overall performance. Still, given that the blocks are small and tasks are short, the gains due to data locality are modest. Even when this is not the best scenario for an active object store such as dataClay, we can see that the gains that dataClay brings outweigh the overall overhead from the distributed storage system.

\begin{figure}
    \centering
    \includegraphics[width=0.95\columnwidth]{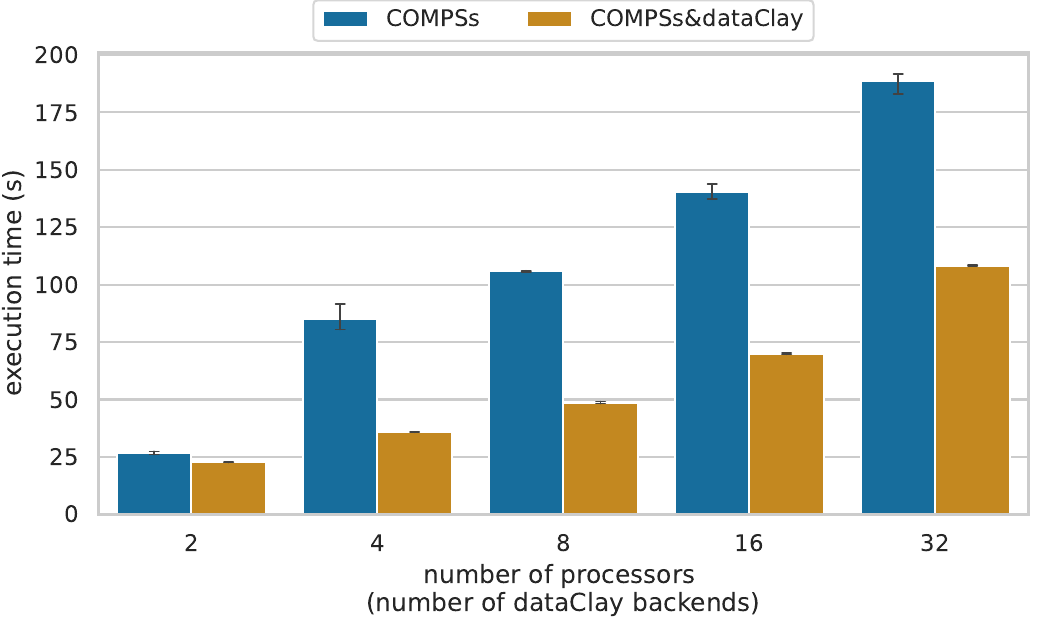}
    \caption{Cascade SVM weak scaling from 2 processors (1 node, 2 backends) to 32 processors (16 nodes, 32 backends), with  24 blocks per processor (one per core)}
    \label{fig:csvm-weakscaling_bigblocks}
\end{figure}

The second set of experiments is shown in Figure~\ref{fig:csvm-weakscaling_bigblocks}. In these experiments we are using the second dataset, the perfectly balanced one in which there is a block per core (or 24 blocks per processor). The dataset size is the same as in the previous experiments, but the size of blocks has increased and the number of tasks has decreased. This means that we are reducing the overhead and general stress upon the distributed framework. The application runs faster than in previous experiments, and the execution is scaling steadily, as expected from the aggregation cost of the Cascade SVM algorithm. Given that the overall number of tasks and intermediate results has decreased, the figure does not show a clear degradation as we could observe in the previous set of experiments.

The advantages of having dataClay increase when the blocks are bigger. This is natural given that the delays and latencies due to data transfers are increased with bigger blocks, and dataClay exploits data locality to avoid this overhead. This reduces bandwidth usage and improves performance and throughput of the distributed system. Overall, we can see significant improvements when using dataClay, that near or exceed the $2\times$ mark in most scenarios.

\section{Reflections and Future Work}
\label{sec:reflections}

This section discusses and elaborates on several aspects of our approach, with the aim of clarifying its current limitations as well as its potential.

While our evaluation used a compact LSTM model for reproducibility, the proposed framework supports models of any complexity through dataClay’s object-oriented interface. The dataClay active storage system handles model complexity transparently, maintaining a consistent programming paradigm whether working with lightweight LSTM networks or resource-intensive architectures such as ResNet-152 or GPT-based models. Its object-level offloading mechanism accommodates demanding workloads, including ResNet-based vision models and GPT-style transformers. Client devices benefit directly from dataClay’s architecture through consistently reduced storage and computational requirements, regardless of underlying model complexity, as all heavy lifting is handled server-side. These advantages become increasingly significant with model size: while our LSTM-based demonstration reduced client storage from 2GB to 0.4GB, modern vision models (EfficientNet, YOLO) require 2–8GB\cite{tan2019efficientnet, terven2023comprehensive}, and Transformer models can range from 10GB to over 70GB\cite{brown2020language, raffel2020exploring}, making storage reduction essential for IoT devices. Memory savings scale similarly: while our LSTM required approximately 330MB (Section~\ref{sec:results}, Tables~\ref{tab:memory_runtime_baseline}–\ref{tab:memory_runtime_dcmacorangepi}), typical CNN inference with ResNet-152 demands 2–8GB\cite{jin2018layer}, and Transformer inference may exceed 16GB depending on sequence length\cite{du2023improving, zhao2024mini}. Most edge devices are incapable of handling such workloads locally, making dataClay’s offloading strategy essential rather than optional.

This scalability is particularly valuable in heterogeneous AI deployments where different edge nodes might trigger various types of inference tasks, from computer vision to natural language processing, while maintaining the same simplified client-side footprint. Furthermore, for heterogeneous AI tasks involving multi-stage inference pipelines, the active storage approach enables efficient chaining of operations. For instance, a typical AI pipeline involving data preprocessing, feature extraction via a CNN, and classification via a transformer can be orchestrated seamlessly across the continuum, with each stage potentially executing on the most suitable available resource while maintaining data locality principles. Our architecture addresses the fundamental "model size vs. device capability" gap in edge AI deployments, with benefits growing proportionally to model complexity.

This article has reviewed the benefits of offloading and it has provided an evaluation with several scenarios, all of them following a \emph{static} offloading mechanism. In real-life deployments, dynamic environments with faulty nodes and faulty connectivity must be considered. Currently, dataClay does have a built-in mechanism for monitoring the health and status of the distributed system (through health-check polling between services). This feature results in automatic failover capabilities (as long as replicas are available), but dataClay does not offer high-level automatic decision-making for replica creation, replica consistency, and load balancing. The design and implementation of strategies that are appropriate for a dynamic continuum environment is left as future work. Those strategies would be determined by evaluating the various parameters and trade-offs (network resiliency, replica overhead, consistency overhead, downtime, bandwidth usage, power efficiency, etc.). The implementation of such strategies could be done as a thin layer --containing all the business logic-- while leveraging the full power of the distributed active object store.

The flexibility of the computing continuum allows to change and upgrade computing devices without redefining the whole continuum architecture. With this in mind, we have performed an evaluation that shows the behavior of different kind of computing devices for the same AI workload; this evaluation shows the impact that the hardware will have on execution times (considering both scenarios: baseline experiments and the dataClay client-server experiments). Network characteristics will have an impact on workloads, and slower networks will be detrimental to any client-server architecture. However, our architecture is flexible and can, for instance, be used in a non-centralized configuration to take advantage of a heterogeneous network. Moreover, a slow network will also result in slower data transfers, increasing the need to optimize data paths; a task offloading solution which includes data management (akin to the solution we propose) is key to leverage unique continuum topologies and make the most of available resources.

Our proposed architecture is perfectly capable of managing soft real-time tasks across the computing continuum. When tasks need to be executed, they can be offloaded and then executed immediately. This fundamental design decision results in a streamlined flow that can support most real-time needs of the application within the continuum. However, a real-life deployment would include additional requirements and capabilities in terms of reliability, latency, redundancy, retries, etc. Such a deployment scenario would require some kind of controller on top of the offloading architecture that we propose in order to satisfy some constraints or metrics --which are application-specific and unique to each scenario.

In the Related Work (Section~\ref{sec:relwork}), we mention Flower \cite{beutel2020flower}, a popular framework for federating ML workloads that is AI-framework agnostic. Just as Flower can leverage existing Python ML frameworks, our proposed architecture is capable of leveraging Flower to achieve a unified solution for data management and task distribution across continuum environments and federated learning scenarios. This integration has been demonstrated in the intelligence layer of the EU-funded ICOS project~\cite{ICOS-paper}, which utilizes multiple clients with training triggered remotely from the intelligence layer and executed in containerized environments that support multiple offloading links per node. The ICOS MetaOS codebase includes OrganizerFL and ModelSync scripts that integrate FL workloads with dataClay\footnote{Available at \url{https://github.com/icos-project/intelligence-module}}. While this federated learning architecture is outside the scope of this article, all the building blocks we have described are compatible with such use cases implemented through Flower. Similarly, as seen in Section~\ref{sec:towards-distributed}, it is possible to use the active storage system with other distributed frameworks (such as PyCOMPSs). This opens the door to generic distributed workloads that can be run across the computing continuum.
\section{Conclusions}
\label{sec:conclusions}

\begin{table*}
    \centering
    \begin{tabular}{|p{2.2cm}|p{6cm}|p{8.5cm}|}
        \hline
        \textbf{Aspect} & \textbf{Benefit} & \textbf{Trade-off / Limitation} \\ \hline \hline
        \textbf{Memory} & Significant reduction of client memory usage. & Higher memory use on server side; aggregated usage (one client + server) increase overall. \\ \hline
        \textbf{Storage} & Client storage shrinks drastically. & Server storage increases. \\ \hline
        \textbf{Execution\newline{}Time} & Faster training/evaluation on powerful devices. & Communication and serialization overhead adds latency, especially on weaker network/devices. \\ \hline
        \textbf{Scalability} & Enables resource pooling and heterogeneous device participation. & Performance depends on network characteristics; high-latency links degrade benefits. \\ \hline
        \textbf{Reliability} & Architecture is capable of supporting replicas and failover mechanisms in dataClay. & No built-in decision-making for replication or load balancing. Implementation would be application-specific. \\ \hline
        \textbf{Programming\newline{}Effort} & Transparent object-oriented interface, low entry barrier for domain experts. & Although it behaves transparently, application code needs to import and use another library (i.e. dataClay). \\ \hline
        \textbf{Deployment} & Easy to containerize. Server side components are already publicly available. & Additional server-side setup and container orchestration. Application specific libraries must be included.\\ \hline
    \end{tabular}
    \caption{Summary of the key benefits and trade-offs when using our proposed architecture.}
    \label{tab:aspects_benefits_tradeoffs}
\end{table*}

This article proposes an innovative way of using active storage systems within computing continuum environments. The architecture we propose is able to offload intelligence workloads across the continuum infrastructure in a transparent way, with a low entry barrier, while maintaining a high level of efficiency.

We have shown an evaluation that compares a baseline (experiments run in a single node, as if they were isolated edge devices) and assessed them in relation to the experiments using the software stack we propose.

As one would expect, the faster execution times were achieved by running powerful devices standalone. However, that is not cost-effective and does not scale; being able to scale (in terms of devices and workloads) and being able to support power-constrained devices is paramount for computing continuum infrastructures. We demonstrate that the architecture we propose gives good support for small devices (responsible for triggering the workloads) while taking advantage of more powerful resources available in the computing continuum, where to store data and offload computation.

In this article, we have also discussed the technical details of a software stack that satisfies the architecture we propose. By using dataClay, a distributed active object store, we were able to perform task offloading on Python applications using mainstream ML libraries. The mechanism to offload computation is based on the object-oriented programming language and comes as a very natural way of programming the application, resulting in a familiar interface that the domain expert can use while retaining a high degree of programmability.

The presence of an active storage system for data management in a computing continuum brings the opportunity to leverage heterogeneous resources available in the infrastructure. The available resources can have different characteristics due to their hardware, their accelerators, etc. Once the architecture contains an active storage system that spans the continuum, task offloading spanning the continuum can be done transparently and reach all those resources.

We include a summary table (Table~\ref{tab:aspects_benefits_tradeoffs}) that highlights the main benefits and costs (trade-offs/limitations) of adopting our active storage–based offloading approach across the computing continuum. This table synthesizes the qualitative aspects presented throughout this article on the following key aspects: memory, storage, execution time, scalability, reliability, programming effort, and deployment.

The evaluation shows how the active storage system enables an infrastructure with small and constrained edge devices that offload tasks to other devices. The software stack that we used for the evaluation is able to achieve this without changing the programming model, agnostic to the libraries being used by the domain expert. Additionally, the distributed evaluation shows that the active storage system 
has the ability to exploit data locality on distributed workloads, bringing a significant speed up to distributed applications.

\section*{Acknowledgements}

This work was partially supported by the projects PID2019-107255GB-C21 and PID2023-147979NB-C21, funded by MCIN / AEI / 10.13039 / 501100011033, and by FEDER, UE.

This work was partially supported by the project ``Towards a
functional continuum operating system (ICOS)'' funded by the European Commission
under Project code/Grant Number 101070177 through the HORIZON EU program.

This work was partially supported by the project ``Open CloudEdgeIoT Platform Uptake in Large Scale Cross-Domain Pilots (O-CEI)'' funded by the European Commission
under Project code/Grant Number 101189589 through the HORIZON EU program.

Anna Queralt is a Serra-Hunter Fellow and has been partially supported by the Spanish Ministerio de Ciencia e Innovación under project / funding scheme PID2023-152841OA-I00 / AEI/10.13039/501100011033 (TALC).


\bibliographystyle{elsarticle-num-names} 
\bibliography{references}

@article{balouek2019towards,
  title={Towards a computing continuum: Enabling edge-to-cloud integration for data-driven workflows},
  author={Balouek-Thomert, Daniel and Renart, Eduard Gibert and Zamani, Ali Reza and Simonet, Anthony and Parashar, Manish},
  journal={The International Journal of High Performance Computing Applications},
  volume={33},
  number={6},
  pages={1159--1174},
  year={2019},
  publisher={SAGE Publications Sage UK: London, England}
}

@incollection{marru2023cybershuttle,
    author = {Marru, Suresh and Pierce, Marlon and Plale, Beth and Pamidighantam, Sudhakar and Wannipurage, Dimuthu and Christie, Marcus and Ranawaka, Isuru and Abeysinghe, Eroma and Quick, Rob and Tajkhorshid, Emad and Koric, Seid and Basney, Jim and Spivak, Mariano and Isralewitz, Barry and Bernardi, Rafael and Gomes, Diego and Krishnan, Giri and Bazhenov, Maxim and Smallen, Shava and Majumdar, Amit and Arkhipov, Anton and Dai, Kael and Liu, Xiao-Ping and Yoshimoto, Kenneth},
    title = {Cybershuttle: An End-to-End Cyberinfrastructure Continuum to Accelerate Discovery in Science and Engineering},
    year = {2023},
    isbn = {9781450399852},
    publisher = {Association for Computing Machinery},
    address = {New York, NY, USA},
    doi = {10.1145/3569951.3593602},
    booktitle = {Practice and Experience in Advanced Research Computing 2023: Computing for the Common Good},
    pages = {26–34},
    numpages = {9},
    location = {Portland, OR, USA},
    series = {PEARC '23}
}

@article{tallent2024final,
  title={Final Report for CHESS: Cloud, High-Performance Computing, and Edge for Science and Security},
  author={Tallent, Nathan and Strube, Jan and Guo, Luanzheng and Lee, Hyungro and Firoz, Jesun and Ghosh, Sayan and Fang, Bo and Bel, Oceane and Spurgeon, Steven and Akers, Sarah and others},
  journal={arXiv preprint arXiv:2410.16093},
  year={2024}
}

@inproceedings{ferrer2021towards,
  title={Towards a cognitive compute continuum: an architecture for ad-hoc self-managed swarms},
  author={Ferrer, Ana Juan and Becker, S{\"o}ren and Schmidt, Florian and Thamsen, Lauritz and Kao, Odej},
  booktitle={2021 IEEE/ACM 21st International Symposium on Cluster, Cloud and Internet Computing (CCGrid)},
  pages={634--641},
  year={2021},
  organization={IEEE}
}

@article{rosendo2022distributed,
  title={Distributed intelligence on the Edge-to-Cloud Continuum: A systematic literature review},
  author={Rosendo, Daniel and Costan, Alexandru and Valduriez, Patrick and Antoniu, Gabriel},
  journal={Journal of Parallel and Distributed Computing},
  volume={166},
  pages={71--94},
  year={2022},
  publisher={Elsevier}
}

@inproceedings{aizman2019high,
  title={High performance I/O for large scale deep learning},
  author={Aizman, Alex and Maltby, Gavin and Breuel, Thomas},
  booktitle={2019 IEEE International Conference on Big Data (Big Data)},
  pages={5965--5967},
  year={2019},
  organization={IEEE}
}

@article{ordonezadaptive,
  title={Adaptive Machine Learning for Resource-Constrained Environments},
  author={Cajas Ordóñez, Sebastián A and Samanta, Jaydeep and {Suárez-Cetrulo, Andrés L.} and Simón Carbajo, Ricardo},
  journal={Discovering Drift Phenomena in Evolving Landscapes},
  pages={3-19},
  publisher={Springer},
  year = {2025}
}

@article{ejarque2022enabling,
  title={{Enabling dynamic and intelligent workflows for HPC, data analytics, and AI convergence}},
  author={Ejarque, Jorge and Badia, Rosa M and Albertin, Lo{\"\i}c and Aloisio, Giovanni and Baglione, Enrico and Becerra, Yolanda and Boschert, Stefan and Berlin, Julian R and D’Anca, Alessandro and Elia, Donatello and others},
  journal={Future generation computer systems},
  volume={134},
  pages={414--429},
  year={2022},
  publisher={Elsevier}
}

@misc{coral,
  title = {Coral},
  howpublished = {\url{https://www.coral.ai/}},
  author = {{Google LLC}},
  note = {Accessed: Jan 2025},
  year = 2020
}

@inproceedings{seshadri2022evaluation,
  title={{An Evaluation of Edge TPU Accelerators for Convolutional Neural Networks}},
  author={Seshadri, Kiran and Akin, Berkin and Laudon, James and Narayanaswami, Ravi and Yazdanbakhsh, Amir},
  booktitle={2022 IEEE International Symposium on Workload Characterization (IISWC)},
  pages={79--91},
  year={2022},
  organization={IEEE}
}

@article{barcelo2023enhancing,
  title={Enhancing iteration performance on distributed task-based workflows},
  author={Barcelo, Alex and Queralt, Anna and Cortes, Toni},
  journal={Future Generation Computer Systems},
  volume={149},
  pages={359--375},
  year={2023},
  publisher={Elsevier}
}

@misc{dataclay-web,
  author = {{BSC - Distributed Objects Management}},
  title = {{dataClay website}},
  howpublished = {\url{https://dataclay.bsc.es}},
  note = {Accessed: March 2025},
  year = 2025
}

@incollection{serrano2021elastic,
  title={An elastic software architecture for extreme-scale big data analytics},
  author={Serrano, Maria A and Mar{\'\i}n, C{\'e}sar A and Queralt, Anna and Cordeiro, Cristovao and Gonzalez, Marco and Pinho, Luis Miguel and Qui{\~n}ones, Eduardo},
  booktitle={Technologies and Applications for Big Data Value},
  pages={89--110},
  year={2021},
  publisher={Springer}
}

@article{bhinder2021artificial,
  title={Artificial intelligence in cancer research and precision medicine},
  author={Bhinder, Bhavneet and Gilvary, Coryandar and Madhukar, Neel S and Elemento, Olivier},
  journal={Cancer discovery},
  volume={11},
  number={4},
  pages={900--915},
  year={2021},
  publisher={American Association for Cancer Research}
}

@article{verma2021artificial,
  title={Artificial intelligence in marketing: Systematic review and future research direction},
  author={Verma, Sanjeev and Sharma, Rohit and Deb, Subhamay and Maitra, Debojit},
  journal={International Journal of Information Management Data Insights},
  volume={1},
  number={1},
  pages={100002},
  year={2021},
  publisher={Elsevier}
}

@article{rong2020artificial,
  title={Artificial intelligence in healthcare: review and prediction case studies},
  author={Rong, Guoguang and Mendez, Arnaldo and Assi, Elie Bou and Zhao, Bo and Sawan, Mohamad},
  journal={Engineering},
  volume={6},
  number={3},
  pages={291--301},
  year={2020},
  publisher={Elsevier}
}

@article{raza2015review,
  title={A review on artificial intelligence based load demand forecasting techniques for smart grid and buildings},
  author={Raza, Muhammad Qamar and Khosravi, Abbas},
  journal={Renewable and Sustainable Energy Reviews},
  volume={50},
  pages={1352--1372},
  year={2015},
  publisher={Elsevier}
}

@article{wang2020convergence,
  title={Convergence of edge computing and deep learning: A comprehensive survey},
  author={Wang, Xiaofei and Han, Yiwen and Leung, Victor CM and Niyato, Dusit and Yan, Xueqiang and Chen, Xu},
  journal={IEEE communications surveys \& tutorials},
  volume={22},
  number={2},
  pages={869--904},
  year={2020},
  publisher={IEEE}
}

@article{soori2023artificial,
  title={Artificial intelligence, machine learning and deep learning in advanced robotics, a review},
  author={Soori, Mohsen and Arezoo, Behrooz and Dastres, Roza},
  journal={Cognitive Robotics},
  volume={3},
  pages={54--70},
  year={2023},
  publisher={Elsevier}
}

@article{shi2016edge,
  title={Edge computing: Vision and challenges},
  author={Shi, Weisong and Cao, Jie and Zhang, Quan and Li, Youhuizi and Xu, Lanyu},
  journal={IEEE internet of things journal},
  volume={3},
  number={5},
  pages={637--646},
  year={2016},
  publisher={Ieee}
}

@article{gatziu1992samos,
  title={{SAMOS: An active object-oriented database system}},
  author={Gatziu, Stella and Dittrich, Klaus R.},
  journal={IEEE Data Eng. Bull.},
  volume={15},
  number={1-4},
  pages={23--26},
  year={1992}
}

@article{marti2017dataclay,
  title={{dataClay: A distributed data store for effective inter-player data sharing}},
  author={Mart{\'i}, Jonathan and Queralt, Anna and Gasull, Daniel and Barcelo, Alex and Costa, Juan Jos{\'e} and Cortes, Toni},
  journal={Journal of Systems and Software},
  volume={131},
  pages={129--145},
  year={2017},
  publisher={Elsevier}
}

@inproceedings{abdelbaky2017computing,
  title={Computing in the continuum: Combining pervasive devices and services to support data-driven applications},
  author={AbdelBaky, Moustafa and Zou, Mengsong and Zamani, Ali Reza and Renart, Eduard and Diaz-Montes, Javier and Parashar, Manish},
  booktitle={2017 IEEE 37th International Conference on Distributed Computing Systems (ICDCS)},
  pages={1815--1824},
  year={2017},
  organization={IEEE}
}

@article{Bisong2019KubeflowPipelines,
    title = {{Kubeflow and Kubeflow Pipelines}},
    year = {2019},
    journal = {Building Machine Learning and Deep Learning Models on Google Cloud Platform},
    author = {Bisong, Ekaba},
    pages = {671--685},
    publisher = {Apress, Berkeley, CA},
    doi = {10.1007/978-1-4842-4470-8_46}
}

@article{Bodor2023MLOps:Directions,
    title = {{MLOps: Overview of Current State and Future Directions}},
    year = {2023},
    journal = {Lecture Notes in Networks and Systems},
    author = {Bodor, Anas and Hnida, Meriem and Najima, Daoudi},
    pages = {156--165},
    volume = {629 LNNS},
    publisher = {Springer Science and Business Media Deutschland GmbH},
    isbn = {9783031268519},
    doi = {10.1007/978-3-031-26852-6_14},
    issn = {23673389}
}

@article{KreuzbergerMachineArchitecture,
  title = {{Machine Learning Operations (MLOps): Overview, Definition, and Architecture}},
  author = {Kreuzberger, Dominik and K{\"{u}}hl, Niklas and Hirschl, Sebastian},
  journal={IEEE access},
  volume={11},
  pages={31866--31879},
  year={2023},
  publisher={IEEE}
}

@article{Zaharia2018AcceleratingMLflow,
    title = {{Accelerating the Machine Learning Lifecycle with MLflow}},
    year = {2018},
    author = {Zaharia, Matei and Chen, Andrew and Davidson, Aaron and Ghodsi, Ali and Hong, Sue Ann and Konwinski, Andy and Murching, Siddharth and Nykodym, Tomas and Ogilvie, Paul and Parkhe, Mani and Xie, Fen and Zumar, Corey},
  journal={IEEE Data Eng. Bull.},
  volume={41},
  number={4},
  pages={39--45}
}

@inproceedings{lofstead2016daos,
  title={{DAOS and Friends: A Proposal for an Exascale Storage System}},
  author={Lofstead, Jay and Jimenez, Ivo and Maltzahn, Carlos and Koziol, Quincey and Bent, John and Barton, Eric},
  booktitle={SC'16: Proceedings of the International Conference for High Performance Computing, Networking, Storage and Analysis},
  pages={585--596},
  year={2016},
  organization={IEEE}
}

@misc{daos-web,
  title = {{DAOS} Administration Guide},
  author = {{Intel Corporation}},
  howpublished = {\url{https://docs.daos.io/}},
  note = {Accessed: March 2025},
  year = 2024
}

@inproceedings{weil2007rados,
  title={{RADOS: A Scalable, Reliable Storage Service for Petabyte-scale Storage Clusters}},
  author={Weil, Sage A and Leung, Andrew W and Brandt, Scott A and Maltzahn, Carlos},
  booktitle={Proceedings of the 2nd international workshop on Petascale data storage: held in conjunction with Supercomputing'07},
  pages={35--44},
  year={2007}
}

@book{fisk2017mastering,
  title={Mastering Ceph},
  author={Nick Fisk},
  publisher={Packt Publishing Ltd},
  year={2017}
}

@misc{openstack-storlets,
  author = {{OpenStack}},
  title = {{Storlets' Documentation}},
  note = {Accessed: March 2025},
  howpublished = {\url{https://docs.openstack.org/storlets/latest/}},
  year = 2024
}

@misc{swift-web,
    author = {{OpenStack}},
    title = {{Swift's Documentation}},
    note = {Accessed: March 2025},
    howpublished = {\url{https://docs.openstack.org/swift/latest/}},
    year = 2023
}

@misc{litert-web,
    author = {{Google AI Edge}},
    title = {{LiteRT}},
    note = {Accessed: March 2025},
    howpublished = {\url{https://ai.google.dev/edge/litert}},
    year = 2025
}

@misc{bentoml-repo,
author = {Yang, Chaoyu and Sheng, Sean and Pham, Aaron and  Zhao, Shenyang and Lee, Sauyon and Jiang, Bo and Dong, Fog and Guan, Xipeng and Ming, Frost},
title = {{BentoML: The framework for building reliable, scalable and cost-efficient AI application}},
howpublished= {\url{https://bentoml.com/}},
note = {Accessed: March 2025},
year = 2025
}

@misc{sagemaker-web,
author = {{Amazon Web Services, Inc.}},
title = {Amazon SageMaker},
howpublished = {\url{https://aws.amazon.com/sagemaker/}},
note = {Accessed: March 2025},
year = 2025
}

@misc{vertexai-web,
    author = {{Google}},
    title = {{Vertex AI}},
    howpublished = {\url{https://cloud.google.com/vertex-ai}},
    note = {Accessed: March 2025},
    year = 2025
}

@misc{s3objectlambda-web,
    author = {{Amazon Web Services, Inc.}},
    title = {{Amazon S3 Object Lambda}},
    howpublished = {\url{https://aws.amazon.com/s3/features/object-lambda/}},
    note = {Accessed: July 2025},
    year = 2025
}

@article{Klaise2020MonitoringProduction,
    title = {{Monitoring and explainability of models in production}},
    year = {2020},
    author = {Klaise, Janis and Van Looveren, Arnaud and Cox, Clive and Vacanti, Giovanni and Coca, Alexandru},
    month = {7},
    url = {https://arxiv.org/abs/2007.06299v1},
    journal={arXiv preprint arXiv:2007.06299},
    arxivId = {2007.06299}
}

@article{Hewage2022MachineSupport,
    title = {{Machine Learning Operations: A Survey on MLOps Tool Support}},
    year = {2022},
    author = {Hewage, Nipuni and Meedeniya, Dulani},
    month = {2},
    doi = {10.48550/arXiv.2202.10169},
    arxivId = {2202.10169v2},
    journal={arXiv preprint arXiv:2202.10169},
}

@article{Chaves2023TheComparative,
    title = {{The orchestration of Machine Learning frameworks with data streams and GPU acceleration in Kafka‐ML: A deep‐learning performance comparative}},
    year = {2023},
    journal = {Expert Systems},
    author = {Chaves, Antonio Jesús and Mart{\'{i}}n, Cristian and D{\'{i}}az, Manuel},
    month = {3},
    doi = {10.1111/exsy.13287},
    issn = {0266-4720}
}

@misc{ray-web,
author = {{Anyscale, Inc.}},
title = {Ray},
howpublished = {\url{https://www.ray.io/}},
note = {Accessed: March 2025},
year = 2025
}

@misc{metaflow-web,
author = {{Netflix, Inc.}},
title = {{Metaflow, A framework for real-life ML, AI, and data science}},
howpublished = {\url{https://metaflow.org/}},
note = {Accessed: March 2025},
year = 2025
}

@misc{openwhisk-web,
author = {{The Apache Software Foundation}},
title = {{Apache OpenWhisk -- Open Source Serverless Cloud Platform}},
howpublished = {\url{https://openwhisk.apache.org/}},
note = {Accessed: July 2025},
year = 2025
}

@misc{openfaas-web,
author = {{OpenFaaS Ltd.}},
title = {{OpenFaaS -- Serverless Functions, Made Simple}},
howpublished = {\url{https://www.openfaas.com/}},
note = {Accessed: July 2025},
year = 2025
}

@misc{knative-web,
author = {{The Knative Authors}},
title = {{Knative}},
howpublished = {\url{https://knative.dev/}},
note = {Accessed: July 2025},
year = 2025
}

@misc{yatai-web,
author = {{BentoML}},
title = {{Yatai: Model Deployment at Scale on Kubernetes}},
howpublished = {\url{https://github.com/bentoml/Yatai}},
note = {Accessed: April 2025},
year = 2024
}

@article{beutel2020flower,
  title={Flower: A Friendly Federated Learning Research Framework},
  author={Beutel, Daniel J and Topal, Taner and Mathur, Akhil and Qiu, Xinchi and Fernandez-Marques, Javier and Gao, Yan and Sani, Lorenzo and Kwing, Hei Li and Parcollet, Titouan and Gusmão, Pedro PB de and Lane, Nicholas D},
  journal={arXiv preprint arXiv:2007.14390},
  year={2020}
}

@article{mukhopadhyay2019heterogeneous,
  title={Heterogeneous integration for artificial intelligence: Challenges and opportunities},
  author={Mukhopadhyay, Saibal and Long, Yun and Mudassar, B and Nair, CS and DeProspo, Bartlet H and Torun, Hakki Mert and Kathaperumal, M and Smet, V and Kim, Duckhwan and Yalamanchili, Sudhakar and others},
  journal={IBM Journal of Research and Development},
  volume={63},
  number={6},
  pages={4--1},
  year={2019},
  publisher={IBM}
}

@article{xu2020edge,
  title={Edge intelligence: Architectures, challenges, and applications},
  author={Xu, Dianlei and Li, Tong and Li, Yong and Su, Xiang and Tarkoma, Sasu and Jiang, Tao and Crowcroft, Jon and Hui, Pan},
  journal={arXiv preprint arXiv:2003.12172},
  year={2020}
}

@article{deng2020edge,
  title={Edge intelligence: The confluence of edge computing and artificial intelligence},
  author={Deng, Shuiguang and Zhao, Hailiang and Fang, Weijia and Yin, Jianwei and Dustdar, Schahram and Zomaya, Albert Y},
  journal={IEEE Internet of Things Journal},
  volume={7},
  number={8},
  pages={7457--7469},
  year={2020},
  publisher={IEEE}
}

@inproceedings{ICOS-paper,
  title = {{ICOS An Intelligent MetaOS for the Continuum}},
  author = {Garcia, Jordi and Masip-Bruin, Xavi and Giannopoulos, Anastasios and Trakadas, Panagiotis and Cajas Ordoñez, Sebastián A. and Samanta, Jaydeep and Suárez-Cetrulo, Andrés L. and Simón Carbajo, Ricardo and Michalke, Marc and Admela, Jukan and Jaworski, Artur and Kotliński, Marcin and Giammatteo, Gabriele and D’Andria, Francesco},
  year = {2025}, 
  isbn = {9798400715600}, 
  publisher = {Association for Computing Machinery}, 
  address = {New York, NY, USA}, 
  doi = {10.1145/3721889.3721929}, 
  booktitle = {Proceedings of the 2nd International Workshop on MetaOS for the Cloud-Edge-IoT Continuum}, 
  pages = {53–59}, 
  numpages = {7}, 
  location = {Rotterdam, Netherlands}, 
  series = {MECC '25} 
}

@article{ray2022review,
  title={{A review on TinyML: State-of-the-art and prospects}},
  author={Ray, Partha Pratim},
  journal={Journal of King Saud University-Computer and Information Sciences},
  volume={34},
  number={4},
  pages={1595--1623},
  year={2022},
  publisher={Elsevier}
}

@article{merlino2024faas,
  title={{FaaS for IoT: Evolving Serverless towards Deviceless in I/Oclouds}},
  author={Merlino, Giovanni and Tricomi, Giuseppe and D’agati, Luca and Benomar, Zakaria and Longo, Francesco and Puliafito, Antonio},
  journal={Future Generation Computer Systems},
  volume={154},
  pages={189--205},
  year={2024},
  publisher={Elsevier}
}

@inproceedings{ali2021hidden,
  title={The hidden cost of the edge: a performance comparison of edge and cloud latencies},
  author={Ali-Eldin, Ahmed and Wang, Bin and Shenoy, Prashant},
  booktitle={Proceedings of the International Conference for High Performance Computing, Networking, Storage and Analysis},
  pages={1--12},
  year={2021}
}

@article{masip2021managing,
  title={Managing the cloud continuum: Lessons learnt from a real fog-to-cloud deployment},
  author={Masip-Bruin, Xavi and Mar{\'\i}n-Tordera, Eva and S{\'a}nchez-L{\'o}pez, Sergi and Garcia, Jordi and Jukan, Admela and Juan Ferrer, Ana and Queralt, Anna and Salis, Antonio and Bartoli, Andrea and Cankar, Matija and others},
  journal={Sensors},
  volume={21},
  number={9},
  pages={2974},
  year={2021},
  publisher={MDPI}
}

@article{jin2018layer,
  title={Layer-centric memory reuse and data migration for extreme-scale deep learning on many-core architectures},
  author={Jin, Hai and Liu, Bo and Jiang, Wenbin and Ma, Yang and Shi, Xuanhua and He, Bingsheng and Zhao, Shaofeng},
  journal={ACM Transactions on Architecture and Code Optimization (TACO)},
  volume={15},
  number={3},
  pages={1--26},
  year={2018},
  publisher={ACM New York, NY, USA}
}

@article{brown2020language,
  title={Language models are few-shot learners},
  author={Brown, Tom and Mann, Benjamin and Ryder, Nick and Subbiah, Melanie and Kaplan, Jared D and Dhariwal, Prafulla and Neelakantan, Arvind and Shyam, Pranav and Sastry, Girish and Askell, Amanda and others},
  journal={Advances in neural information processing systems},
  volume={33},
  pages={1877--1901},
  year={2020}
}

@article{raffel2020exploring,
  title={Exploring the limits of transfer learning with a unified text-to-text transformer},
  author={Raffel, Colin and Shazeer, Noam and Roberts, Adam and Lee, Katherine and Narang, Sharan and Matena, Michael and Zhou, Yanqi and Li, Wei and Liu, Peter J},
  journal={Journal of machine learning research},
  volume={21},
  number={140},
  pages={1--67},
  year={2020}
}

@inproceedings{tan2019efficientnet,
  title={Efficientnet: Rethinking model scaling for convolutional neural networks},
  author={Tan, Mingxing and Le, Quoc},
  booktitle={International conference on machine learning},
  pages={6105--6114},
  year={2019},
  organization={PMLR}
}

@article{terven2023comprehensive,
  title={A comprehensive review of yolo architectures in computer vision: From yolov1 to yolov8 and yolo-nas},
  author={Terven, Juan and C{\'o}rdova-Esparza, Diana-Margarita and Romero-Gonz{\'a}lez, Julio-Alejandro},
  journal={Machine learning and knowledge extraction},
  volume={5},
  number={4},
  pages={1680--1716},
  year={2023},
  publisher={MDPI}
}

@article{du2023improving,
  title={Improving computation and memory efficiency for real-world transformer inference on gpus},
  author={Du, Jiangsu and Jiang, Jiazhi and Zheng, Jiang and Zhang, Hongbin and Huang, Dan and Lu, Yutong},
  journal={ACM Transactions on Architecture and Code Optimization},
  volume={20},
  number={4},
  pages={1--22},
  year={2023},
  publisher={ACM New York, NY}
}

@article{zhao2024mini,
  title={Mini-Sequence Transformers: Optimizing Intermediate Memory for Long Sequences Training},
  author={Zhao, Jiawei and Chen, Zhuoming and Chen, Beidi and Anandkumar, Animashree and others},
  journal={Advances in Neural Information Processing Systems},
  volume={37},
  pages={97299--97327},
  year={2024}
}

@inproceedings{kale1993charm++,
  title={{Charm++ A Portable Concurrent Object Oriented System Based On C++}},
  author={Kale, Laxmikant V and Krishnan, Sanjeev},
  booktitle={Proceedings of the eighth annual conference on Object-oriented programming systems, languages, and applications},
  pages={91--108},
  year={1993}
}

@inproceedings{galvez2018charmpy,
  title={{CharmPy: A Python Parallel Programming Model}},
  author={Galvez, Juan J and Senthil, Karthik and Kale, Laxmikant},
  booktitle={2018 IEEE International Conference on Cluster Computing (CLUSTER)},
  pages={423--433},
  year={2018},
  organization={IEEE}
}

@misc{zodb-web,
author = {{Zope Foundation}},
title = {{ZODB - a native object database for Python}},
howpublished = {\url{https://zodb-docs.readthedocs.io/}},
note = {Accessed: October 2025},
year = 2025
}

@article{graf2004parallel,
  title={Parallel support vector machines: The cascade SVM},
  author={Graf, Hans and Cosatto, Eric and Bottou, Leon and Dourdanovic, Igor and Vapnik, Vladimir},
  journal={Advances in neural information processing systems},
  volume={17},
  year={2004}
}

@inproceedings{dislib,
            title       = {{dislib: Large Scale High Performance Machine Learning in Python}},
            author      = {Javier Álvarez Cid-Fuentes and Salvi Solà and Pol Álvarez and Alfred Castro-Ginard and Rosa M. Badia},
            booktitle   = {Proceedings of the 15th International Conference on eScience},
            pages       = {96-105},
            year        = {2019},
 }

@article{tejedor2017pycompss,
  title={PyCOMPSs: Parallel computational workflows in Python},
  author={Tejedor, Enric and Becerra, Yolanda and Alomar, Guillem and Queralt, Anna and Badia, Rosa M and Torres, Jordi and Cortes, Toni and Labarta, Jes{\'u}s},
  journal={The International Journal of High Performance Computing Applications},
  volume={31},
  number={1},
  pages={66--82},
  year={2017},
  publisher={Sage Publications Sage UK: London, England}
}

@misc{marenostrum,
author = {{Barecelona Supercomputing Center}},
title = {{MareNostrum}},
month = {December},
year = {2022},
howpublished={\url{https://www.bsc.es/marenostrum/marenostrum}}
}

\end{document}